\pdfoutput=1
\documentclass[11pt,twoside,a4paper,cmspaper,final,collab]{cms-tdr}
\def\svnVersion{0b9878b}\def\svnDate{2025/10/31}\def\cmsCernNoTag{CERN-EP-2025-017}\def\cmsCernDate{\today}\def\cmsMessage{Published in the Journal of High Energy Physics as \href{http://dx.doi.org/10.1007/JHEP10(2025)236}{\doi{10.1007/JHEP10(2025)236}.}}
\begin{document}\cmsNoteHeader{SUS-23-001}

\newlength\cmsTabSkip\setlength{\cmsTabSkip}{1ex}
\providecommand{\cmsLeft}{left\xspace}
\providecommand{\cmsRight}{right\xspace}
\providecommand{\cmsTable}[1]{\resizebox{\textwidth}{!}{#1}}
\newcommand{\snn}  {\ensuremath{S_\mathrm{NN}}\xspace}
\newcommand{\snni}[1]  {\ensuremath{S_\mathrm{NN,#1}}\xspace}
\newcommand{\sgo}{\ensuremath{\widetilde{\mathrm{S}}}\xspace}
\newcommand{\lsp}{\PSGczDo}
\newcommand{\singlet}{\HepParticle{S}{}{}\xspace}
\newcommand{\mstop}{\ensuremath{m_{\PSQt}}\xspace}
\newcommand{\syy}  {\ensuremath{\singlet \HepParticle{Y}{}{} \HepAntiParticle{Y}{}{} }\xspace}

\newcommand{\ttjets}{\ensuremath{\ttbar + \text{jets}}\xspace}
\newcommand{\QCDjets}{\ensuremath{\text{QCD multijet}}\xspace}

\newcommand{\njets}{\ensuremath{N_\text{jets}}\xspace}
\newcommand{\nbjets}{\ensuremath{N_{\cPqb}}\xspace}
\newcommand{\ntops}{\ensuremath{N_\PQt}\xspace}
\newcommand{\dRbjets}{\ensuremath{\Delta R_{\cPqb \text{jets}}}\xspace}
\newcommand{\mbl}{\ensuremath{M_{\cPqb\ell}}\xspace}
\newcommand{\mll}{\ensuremath{M_{\ell \ell}}\xspace}
\newcommand{\signalxsec}{\ensuremath{\sigma_{\sTop\overline{\sTop}}}\xspace}
\newcommand{\nonclosure}{\ensuremath{{C}\hspace{-0.70em}/\kern 0.0em}\xspace}

\newcommand{\Areg}{\ensuremath{A}\xspace}
\newcommand{\Breg}{\ensuremath{B}\xspace}
\newcommand{\Creg}{\ensuremath{C}\xspace}
\newcommand{\Dreg}{\ensuremath{D}\xspace}
\newcommand{\NAreg}{\ensuremath{N_{A}}\xspace}
\newcommand{\NApred}{\ensuremath{N_{A,\;\text{pred.}}\xspace}}
\newcommand{\NBreg}{\ensuremath{N_{B}}\xspace}
\newcommand{\NCreg}{\ensuremath{N_{C}}\xspace}
\newcommand{\NDreg}{\ensuremath{N_{D}}\xspace}

\newcommand{\Fig}[1]{Fig.~\ref{fig:#1}}
\newcommand{\Tab}[1]{Table~\ref{tab:#1}} 
\newcommand{\Equ}[1]{Eq.~(\ref{eq:#1})}
\newcommand{\Sec}[1]{Section~\ref{sec:#1}}

\newcommand{\Figure}[1]{Figure~\ref{fig:#1}}
\newcommand{\Table}[1]{Table~\ref{tab:#1}}

\newcommand{\discotec}{ABCDisCoTEC\xspace}

\newcommand{\zerol}{0$\Pell$\xspace}
\newcommand{\onel}{1$\Pell$\xspace}
\newcommand{\twol}{2$\Pell$\xspace}

\cmsNoteHeader{SUS-23-001}

\title{Search for top squarks in final states with many light-flavor jets and 0, 1, or 2 charged leptons in proton-proton collisions at \texorpdfstring{$\sqrt{s}=13\TeV$}{sqrt(s) = 13 TeV}}

\date{\today}

\abstract{
Several new physics models including versions of supersymmetry (SUSY) characterized by $R$-parity violation (RPV) or with additional hidden sectors predict the production of events with top quarks, low missing transverse momentum, and many additional quarks or gluons. The results of a search for top squarks decaying to two top quarks and six additional light-flavor quarks or gluons are reported. The search employs a novel machine learning method for background estimation from control samples in data using decorrelated discriminators. The search is performed using events with 0, 1, or 2 electrons or muons in conjunction with at least six jets. No requirement is placed on the magnitude of the missing transverse momentum. The result is based on a sample of proton-proton collisions at $\sqrt{s} = 13\TeV$ corresponding to 138\fbinv of integrated luminosity collected with the CMS detector at the LHC in 2016--2018. With no statistically significant excess of events observed beyond the expected contributions from the standard model, the data are used to determine upper limits on the top squark pair production cross section in the frameworks of RPV and stealth SUSY. Models with top squark masses less than 700 (930)\GeV are excluded at 95\% confidence level for RPV (stealth) SUSY scenarios.
}

\hypersetup{
pdfauthor={CMS Collaboration},
pdftitle={Search for top squarks in final states with many light-flavor jets and 0, 1, or 2 charged leptons in proton-proton collisions at sqrt(s) = 13 TeV},
pdfsubject={CMS},
pdfkeywords={CMS, supersymmetry, SUSY}}

\maketitle 

\section{Introduction}
\label{sec:intro}

The standard model (SM) of particle physics describes observed phenomena of
elementary particles with incredible precision.
Despite this success, there remain many questions unanswered
by the SM that inspire a continuing
search for new particles and phenomena predicted by theories beyond the SM (BSM).
Supersymmetry (SUSY) is a long-sought-after BSM framework that addresses
several shortcomings of the SM including the gauge hierarchy problem,
in which radiative corrections to the Higgs boson mass parameter,
especially those associated with the top quark, are quadratically divergent.
In SUSY, this hierarchy problem can be mitigated by the presence of a light
bosonic top squark \PSQt, which, as the supersymmetric partner of the top quark,
brings an additional set of radiative corrections that largely cancel the
divergent SM corrections~\cite{susy, Martin:1997ns}.
Historically, searches for strong production of \PSQt squarks have focused
on final states with large missing transverse momentum \ptmiss related to the
undetected lightest SUSY particle (LSP) produced in \PSQt squark decays; strong
limits have been placed on \PSQt squark production in this context~\cite{CMS:2019ysk, CMS:2019zmd, CMS:2019ybf, CMS:2021beq, ATLAS:2020dsf, ATLAS:2020xzu, ATLAS:2021hza, ATLAS:2024rcx}.
However, well-motivated variants of SUSY can produce final states
without large \ptmiss including SUSY with $R$-parity violation
(RPV)~\cite{rpv, Dreiner:1991pe} and ``stealth" SUSY~\cite{stealthSusy1, Fan:2012jf, Fan:2015mxp}.

\begin{figure}[htpb]
    \centering
    \includegraphics[width=0.4\textwidth]{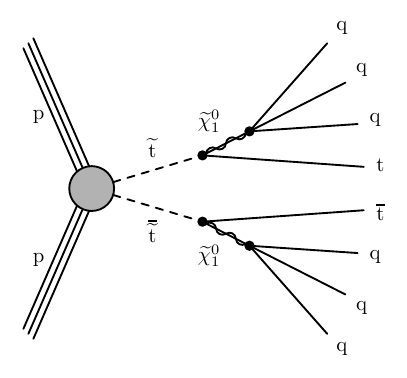} 
    \includegraphics[width=0.4\textwidth]{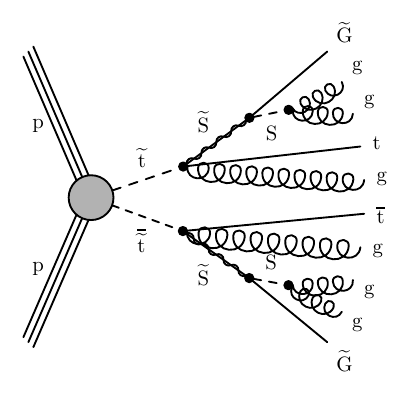}
    \caption{ Diagrams of top squark pair production with each squark decaying to a top quark and additional 
    light-flavor quarks for the RPV SUSY model (\cmsLeft) and with each squark decaying to a top quark,
    gluons, and a gravitino for the stealth \syy model (\cmsRight).
    \label{fig:feynmanDiagrams}}
 \end{figure}

In this analysis, we search for \PSQt squark pair production with decays in both
the RPV and stealth SUSY frameworks.
In RPV SUSY, additional superpotential terms
allow the LSP to decay into SM particles, thereby producing a signature with
balanced transverse momentum \pt.
The RPV coupling governing hadronic decays of SUSY particles is denoted 
by $\Lampp_{abc}$ where $a,b,c$ indicate the generation of the one up-type and two down-type 
quarks involved in the decay vertex~\cite{rpv}. 
In the benchmark RPV model considered here, each \PSQt squark decays to a top
quark \PQt and a neutralino \lsp, followed by the RPV decay of the \lsp to three
light-flavor quarks, $\lsp\to\PQq\PQq\PQq$.
Since the analysis is insensitive to differences between jets originating from
quarks of the first and second generations, the results are applicable to 
RPV models with coupling $\Lampp_{abc}$ where $a,b,c \in \{1, 2\}$.
The stealth SUSY \syy model~\cite{Fan:2012jf} assumes a
new hidden sector containing a scalar particle \singlet with even $R$-parity 
and its superpartner \sgo, both of which are singlets under all SM interactions~\cite{stealthSusy1}.
The key feature of stealth SUSY is that supersymmetry is approximately conserved in the
hidden sector, resulting in nearly mass-degenerate \sgo and \singlet without
artificial tuning of particle masses.  In the benchmark \syy model,
each \PSQt squark decays to a top quark, gluon $\Pg$, and a singlino $\sgo$, with the $\sgo$ decay contributing
a gravitino $\PXXSG$ and two additional gluons to the final state, 
as shown in Fig.~\ref{fig:feynmanDiagrams}.
Because of the small \singlet--\sgo mass difference and small \PXXSG mass,
the undetected \PXXSG results in negligible observed \ptmiss.

The CMS Collaboration at the CERN LHC published a search for these signals in 2021~\cite{SUS-19-004}
based on the full Run 2 data set that considered final states with
exactly one lepton and at least seven jets.
The results showed a maximum local signal significance of 2.8 standard deviations
for a \PSQt squark mass \mstop of 400\GeV in the RPV model, arising primarily from
terms in the fit likelihood for
constrained nuisance parameters
associated with the systematic uncertainty in the shape of the
jet multiplicity (\njets) distribution for
the dominant \ttjets background.

The search presented in this paper extends and improves on the
previous analysis in two ways.

First, to better evaluate whether any observed discrepancy arises in events
with two top quarks in the final state, we now consider all three
$\ttbar$ decay channels: all-hadronic (\zerol),
single-lepton (\onel), and fully leptonic (\twol).
This results in a ${\approx}30\%$ improvement in the expected upper limit on the \PSQt squark pair production cross section
for the RPV signal model with $\mstop=850\GeV$ (near the previous exclusion limit on \mstop).

Second, the method for estimating the primary \ttjets background
from data has been updated to reduce the dependence of the analysis on
systematic uncertainties related to the \njets modeling in \ttjets events~\cite{SUS-19-004}.
In the previous analysis, the maximum likelihood fit, which provides a simultaneous estimate of the \ttjets background and signal,
relied on modeling of the shape of the \njets distribution for the background. 
In the current analysis, however, the fit employs independent background yield parameters for each \njets category.  
These yield parameters are constrained using an effective sideband extrapolation in two independent neural network (NN)
variables that discriminate signal from the background (called \snni{1} and \snni{2}), which are generated according to the ABCDisCo Training Enhanced with Closure (ABCDisCoTEC) method~\cite{CMS:2025cvw}. 
This method augments the ABCDisCo technique~\cite{ABCDmethod_2021}, which minimizes the distance correlation between \snni{1} and \snni{2}, by also directly minimizing the nonclosure of the sideband extrapolation in the NN training.
This technique yields an overall reduction in systematic uncertainty arising from possible discrepancies in the shape of the \njets spectrum.
The expected upper limit on the \PSQt squark pair production cross section has improved by a factor of ${\approx}2$ at low \mstop (between 300 and 450\GeV) for both signal models as a result of this improvement.

The detector and general experimental techniques are
described in Section~\ref{sec:experimental}, followed by a discussion of
the simulated samples in Section~\ref{sec:sim}. 
The event selection is detailed in Section~\ref{sec:sele}.
Section~\ref{sec:discoTEC} outlines the \discotec method and its application in this analysis. 
The background estimation and treatment of systematic uncertainties are detailed
in Section~\ref{sec:bkg} and \ref{sec:syst}, respectively.
Finally, the results and their interpretation are described
in Section~\ref{sec:res}, followed by a summary in Section~\ref{sec:sum}.
Tabulated results are provided in the HEPData record for this analysis~\cite{hepdata}.

\section{Experimental techniques}
\label{sec:experimental}

This search examines proton-proton ($\Pp\Pp$) collision data at $\sqrt{s} = 13\TeV$ collected during 2016--2018 with the CMS detector at the LHC corresponding to an integrated luminosity of 138\fbinv \cite{CMS:LUM-17-003, CMS:LUM-17-004, CMS:LUM-18-002}.
The CMS apparatus~\cite{CMS:2008xjf,CMS:2023gfb} is a multipurpose, nearly hermetic detector, designed to trigger on~\cite{CMS:2020cmk,CMS:2016ngn,CMS:2024aqx} and identify electrons, muons, photons, and (charged and neutral) hadrons \cite{CMS:2020uim,CMS:2018rym,CMS:2014pgm}.
Its central feature is a superconducting solenoid of 6\unit{m} internal diameter, providing a magnetic field of 3.8\unit{T}.
Within the solenoid volume are a silicon pixel and strip tracker, a lead tungstate crystal electromagnetic calorimeter (ECAL), and a brass and scintillator hadron calorimeter (HCAL), each composed of a barrel and two endcap sections. 
Forward calorimeters extend the pseudorapidity coverage provided by the barrel and endcap detectors.
Muons are reconstructed in gas-ionization detectors embedded in the steel flux-return yoke outside the solenoid. 
More detailed descriptions of the CMS detector, together with a definition of the coordinate system used and the relevant kinematic variables, can be found in Refs.~\cite{CMS:2008xjf,CMS:2023gfb}.

The primary vertex (PV) is taken to be the vertex corresponding to the hardest scattering in the event, evaluated using tracking information alone, as described in Section 9.4.1 of Ref.~\cite{CMS-TDR-15-02}.
A particle-flow algorithm~\cite{CMS:2017yfk} aims to reconstruct and identify each individual particle in an event, with an optimized combination of information from the various elements of the CMS detector. 
The energy of photons is obtained from the ECAL measurement. 
The energy of electrons is determined from a combination of the electron momentum at the PV as determined by the tracker, the energy of the corresponding ECAL cluster, and the energy sum of all bremsstrahlung photons spatially compatible with originating from the electron track. 
The energy of muons is obtained from the curvature of the corresponding track. 
The energy of charged hadrons is determined from a combination of their momentum measured in the tracker and the matching ECAL and HCAL energy deposits, corrected for the response function of the calorimeters to hadronic showers. 
Finally, the energy of neutral hadrons is obtained from the corresponding corrected ECAL and HCAL energies.

For each event, hadronic jets are clustered from these reconstructed particles using the anti-\kt algorithm~\cite{Cacciari:2008gp, Cacciari:2011ma} with a distance parameter of 0.4.
Jet momentum is determined as the vectorial sum of all particle momenta in the jet, and is found from simulation to be, on average, within 5 to 10\% of the true momentum over the whole \pt spectrum and detector acceptance.
Additional $\Pp\Pp$ interactions within the same or nearby bunch crossings (pileup) can contribute additional tracks and calorimetric energy depositions to the jet momentum. To mitigate this effect, charged particles identified to be originating from pileup vertices are discarded and an offset correction is applied to correct for remaining contributions~\cite{CMS:2020ebo}.
Jet energy corrections are derived from simulation studies so that the average measured energy of jets becomes identical to that of particle level jets. 
In situ measurements of the momentum balance in dijet, $\text{photon} + \text{jet}$, $\PZ + \text{jet}$, and multijet events are used to determine any residual differences between the jet energy scale in data and in simulation, and appropriate corrections are made~\cite{CMS:2016lmd}. 
Additional selection criteria are applied to each jet to remove jets potentially dominated by instrumental effects or reconstruction failures~\cite{CMS:2020ebo}.

For this analysis, reconstructed jets are required to have $\pt > 30\GeV$ and $|\eta| < 2.4$.
Jets that overlap with a lepton meeting the analysis requirements within a cone of radius $\Delta R = \sqrt{\smash[b]{(\Delta \phi)^2 + (\Delta \eta)^2}} = 0.4$ are removed.
These jets are evaluated using the ``medium'' working point of the DeepJet tagging algorithm designed to identify jets as arising from bottom quark fragmentation (\cPqb-tagged jet).
For jets with $\pt \approx 30\GeV$, the efficiency of the algorithm is ${\approx}75\%$ with a misidentification rate for light-flavor jets and gluon jets of 1\%~\cite{bTagging,deepJet1,deepJet2}. 

Electrons are required to pass a ``tight'' level of identification that is around 70\% efficient in electron identification \cite{CMS:2020uim,CMS-DP-2020-021}.
All electrons in this analysis are required to meet the requirements of $\pt > 30\GeV$ and $|\eta| < 2.4$. 
For data in 2017--2018, the \pt threshold is raised to 37\GeV to accommodate higher trigger thresholds. 
The transverse and longitudinal impact parameters of the electron track from the PV are required to be less than 0.05 (0.10)\unit{cm} in the barrel (endcap) and less than 0.10 (0.20)\unit{cm} in the barrel (endcap), respectively.

Muons are required to pass a ``medium'' level of identification that is around 98\% efficient for genuine muon identification \cite{CMS:2018rym}. 
Muons are additionally required to have $\pt > 30\GeV$ and $|\eta| < 2.4$.
The transverse (longitudinal) impact parameters of the muon track from the PV are required to be less than 0.2 (0.5)\unit{cm}, respectively.

Additionally, electrons and muons are required to be isolated from nearby detector activity, 
as defined by the requirement that the ratio of the \pt sum of other objects in a cone surrounding the lepton to the \pt of the lepton itself be less than 0.1 (0.2) for electrons (muons), where the radius of the cone scales as $1 / \pt$~\cite{miniIso} and ranges from a minimum of 0.05 for leptons with $\pt > 200\GeV$ to a maximum of 0.2 for $\pt < 50\GeV$.

In addition to the muons defined above, a second collection of muons is used to define a control region (CR) for multijet production from quantum chromodynamics (QCD multijet). 
These muons, called nonisolated muons, have the same ``medium'' identification, $\eta$, and impact parameter requirements as standard muons, but they are required to have $\pt > 55\GeV$ and the isolation requirement is reversed. 

Events of interest are selected using a two-tiered trigger system.
The first level (L1), composed of custom hardware processors, uses information from the calorimeters and muon detectors to select events at a rate of around 100\unit{kHz} within a fixed latency of 4\mus~\cite{CMS:2020cmk}.
The second level, known as the high-level trigger, consists of a farm of processors running a version of the full event reconstruction software optimized for fast processing, and reduces the event rate to around 1\unit{kHz} before data storage~\cite{CMS:2016ngn}.

For the \zerol channel, a collection of triggers that target large amounts of hadronic activity and \cPqb-tagged jets are employed. 
The most efficient of these triggers requires six jets with $\pt > 40\GeV$, a \PQb-tagged jet, and the scalar \pt sum of all jets, \HT, to be above 450 (430)\GeV, for data collected in 2016 (2017--2018). 
Additional hadronic triggers are also included that require events to have at least one jet with $\pt > 450\,(500)\GeV$ or $\HT > 900\,(1050)\GeV$, with no other restrictions, for data collected in 2016 (2017--2018).
In 2016, this combination of triggers achieves an overall efficiency of 85--90\% near the thresholds of $\HT > 450\GeV$ or $\pt > 40\GeV$ for the sixth highest ranked \pt jet.
In 2017--2018, among events with only one \PQb-tagged jet (at least 2 \PQb-tagged jets), the overall efficiency is about 55 (80)\% near the thresholds of $\HT > 500\GeV$ or sixth jet $\pt > 40\GeV$.

In both the \onel and \twol channels, events are selected with triggers requiring at least one electron or one muon.
The lowest \pt thresholds for these triggers in 2016 (2017--2018) are 27 (35)\GeV for electrons and 24 (24)\GeV for muons. 
Additionally, the leptons are required to be isolated from tracks and from calorimeter activity. 
Lepton triggers without isolation requirements are also used with higher thresholds of 115\GeV for electrons and 50\GeV for muons.
All leptonic triggers are combined for an overall efficiency ranging from about 70\% for leptons with \pt near the trigger threshold to above 85\% for $\pt > 120\GeV$.
The single-muon trigger with the 50\GeV \pt threshold is additionally used to populate the QCD multijet CR. 
The efficiency of this trigger alone is about 90\% for muons with $\pt > 100\GeV$.

During the 2016 and 2017 data taking, a gradual shift in the timing of the inputs of the ECAL L1 trigger in the region at $\abs{\eta} > 2.0$ caused a specific trigger inefficiency. 
For events containing an electron (a jet) with \pt larger than $\approx$50\GeV ($\approx$100\GeV), in the region $2.5 < \abs{\eta} < 3.0$, the efficiency loss is $\approx$10--20\%, depending on \pt, $\eta$, and time.

\section{Simulated samples}
\label{sec:sim}

Simulated samples are used to estimate contributions from minor backgrounds to the search region (SR) yields, 
to model the signal contribution in the SR, to estimate systematic uncertainties, 
and to serve as inputs (signal and \ttbar) to the training of NNs.

Top quark pair production, $t$-channel single top quark production,
and single top quarks produced in association with a \PW boson 
are generated with the next-to-leading order (NLO) \POWHEG v2.0 generator~\cite{powheg1,powheg2,powheg3,powhegttbar,powhegtW,powhegsingletopschannel}.
The \MGvATNLO v2.6.5 generator~\cite{madgraph} is used in NLO mode for 
$\ttbar + \PW/\PZ/\PH$, $\ttbar\ttbar$, $\PQt\PZ\Pq$, and $s$-channel single top quark production events,
while leading order (LO) mode is used for QCD multijet, \ttbar + diboson/$\PQt\PW$/$\PQt + \text{jet}$, Drell--Yan, and $\PW + \text{jets}$ production.
Diboson production is generated at LO with \PYTHIA v8.240~\cite{pythia}.
Samples are normalized to production cross sections calculated to the highest available accuracy, generally NLO or next-to-NLO (NNLO)~\cite{madgraph,Czakon:2011xx,Kant:2014oha,Aliev:2010zk,Gehrmann:2014fva,Campbell:1999ah,Campbell:2011bn,Li:2012wna}.

For the signal, top squarks are produced with \MGvATNLO in LO mode with up to two
additional partons at the matrix element level and their decays are simulated with \PYTHIA. 
Signal production cross sections (\signalxsec) are approximated to NNLO plus
next-to-next-to-leading logarithmic (NNLL) accuracy~\cite{signalxsecNNLO,signalxsecNNLL,Beenakker:2016gmf,Beenakker:2010nq,Beenakker:1997ut}.
A range of \mstop is considered from 300 to 1050\GeV, and all decays are assumed
to be prompt with branching fractions of unity.
In the benchmark RPV model, we take the \lsp mass to be 100\GeV, which
ensures an on-shell top quark in the \PSQt squark decay for all \mstop.
For the benchmark stealth \syy model, the characteristic
small \singlet--\sgo mass difference is held constant at 10\GeV, 
and we assume an \sgo mass of 100\GeV and a \PXXSG mass of 1\GeV.  

Hadronization and parton showering is done by \PYTHIA using underlying event (UE) tune CP5~\cite{Sirunyan_2020,TuneCP5}.
Parton distribution functions (PDFs) from the NNPDF3.1 NNLO set~\cite{NNPDF} are used for all processes.
The CMS detector response is modeled in simulation using \GEANTfour~\cite{GEANT}.
This response is then passed through the same event reconstruction as collision data.

In order to match the instantaneous luminosity profile measured at the CMS detector, simulated events are reweighted based on the distributions of the observed and simulated number of $\Pp\Pp$ interactions per bunch crossing.
Corrections related to the efficiencies of the trigger system (including for the ECAL timing shift),
lepton identification and isolation (for the \onel and \twol channels), \PQb tagging, and top quark tagging (for the \zerol channel)
are applied to simulated events in order to mitigate small differences observed with respect to data.
These corrections are calculated by comparing the efficiency of object identification in simulation to that in data.

\section{Event selection}
\label{sec:sele} 

The dominant background for the search arises from
top quark pair production with additional jets from
initial- and final-state radiation (ISR and FSR).
The primary physical
observable that allows discrimination of signal from \ttjets
background is \njets, which is enhanced in the signal.
QCD multijet production is 
the subdominant background with additional minor backgrounds from
production of \ttbar in association with SM weak gauge bosons or
additional top quarks ({\ttbar}+X), production of weak gauge bosons,
and single top quark production---all with additional jets
from ISR or FSR.
The QCD multijet and other non-\ttbar backgrounds are suppressed via selection
criteria related to the presence of a top quark. We require either
one or two charged leptons (\Pe or \Pgm) or, for events with zero charged leptons, the identification
of a top quark via a top quark tagging algorithm. 
Further suppression is obtained with requirements based on the
presence of \cPqb-tagged jets. 

The selection criteria for the three search channels are summarized in \Tab{signalRegions}.
The three search region selections are defined orthogonally using exact selections on the number of isolated leptons $N_{\text{leptons}}^{\text{iso}}$.
Each selection is defined such that it is more restrictive than the trigger requirements for jets, leptons, and $\HT$ for the respective channels. 
The particular \njets selection for each channel follows from the leptonic versus hadronic decay of each \PW boson, while also considering effects from detector acceptance and contributions from other backgrounds. 
Additional restrictions are placed on the jets for the \zerol channel to match the jet \pt requirements in the trigger. 
Jets passing these additional restrictions are a subset of the jets that pass the less strict \pt requirement.

Each of the three search channels is also defined with dedicated selections to enhance background rejection based on the unique final-state topology of each channel. 
The \zerol channel selection requires at least two \PQb-tagged jets (\nbjets), which are angularly separated by $\dRbjets > 1$ to minimize the QCD multijet event yield.
This background is further reduced by requiring at least two tagged top quark objects (\ntops) identified by an NN-based hadronic top quark tagging algorithm~\cite{mergedTopTagging, CMS:2021beq}. 
The \PQt tagging algorithm employs two different NNs, with one network trained to identify low-\pt top quarks (${\lesssim} 400\GeV$) that decay to three individually resolved jets, and 
the other to identify high-\pt top quarks from single jets clustered with the anti-\kt algorithm with a distance parameter of 0.8.
The working point for the \PQt tagging algorithm is optimized in order to maximize statistical significance in the \zerol search channel.
The average efficiency for the two taggers is ${\approx} 40\%$. 

The \onel channel event selection imposes that the invariant mass of the required \PQb-tagged jet and lepton (\mbl) be between 50 and 250\GeV to loosely tag the leptonically decaying top quark in the event.
Finally, the \twol selection requires that the invariant mass of the two oppositely charged leptons (\mll) be distant from the \PZ boson mass resonance peak in order to reduce the  Drell--Yan background. 

\begin{table}[t]
    \centering
    \topcaption{Search region selections for the \zerol, \onel, and \twol channels. The term $N_{\text{muon}}^{\text{noniso}}$ denotes the number of nonisolated muons. 
    \label{tab:signalRegions}}
    \def\arraystretch{1.2}
    \begin{tabular}{l l c c c}
      \multicolumn{2}{l}{Selection criteria}                 & \zerol     &  \onel       & \twol \\
      \hline                                                          
      \multicolumn{2}{l}{$N_{\text{leptons}}^{\text{iso}}$}  & 0          &  1           & 2, oppositely charged \\
      \multicolumn{2}{l}{$N_{\text{muon}}^{\text{noniso}}$} & 0          &  0           & 0                     \\
      \multicolumn{2}{l}{\HT (\GeVns)}                       & $>$500    &  $>$500     & $>$500               \\
      \njets & $(\pt > 30\GeV)$ & ${\geq}8$ & ${\geq}7$ & ${\geq}6$\\
      \njets & $(\pt > 45\GeV)$ & ${\geq}6$ & \NA & \NA \\
      \nbjets & $(\pt > 30\GeV)$ & ${\geq}2$ & ${\geq}1$ & ${\geq}1$\\
      \nbjets & $(\pt > 45\GeV)$ & ${\geq}1$ & \NA & \NA \\
      \multicolumn{2}{l}{$\ntops$}                           & $\geq$2  & \NA          & \NA                     \\
      \multicolumn{2}{l}{$\mbl$ (\GeVns) }                     & \NA        & $>$50, $<$250  & \NA                     \\
      \multicolumn{2}{l}{$\mll$ (\GeVns) }                     & \NA        & \NA          & $<$81 or $>$101     \\
      \multicolumn{2}{l}{$\dRbjets$}         & $>$1   & \NA          & \NA                     \\        
    \end{tabular}
\end{table}

For each of the three analysis channels, a unique QCD multijet CR is used to estimate the QCD multijet event yield in the SR. 
These events are identified by the presence of a nonisolated muon along with the same requirements for \njets and \HT used for the corresponding SR definitions.
The requirement of exactly one nonisolated muon ensures that the CR is sufficiently pure in multijet events and matches the trigger requirements.
No additional requirements are placed on \nbjets or \ntops.

Simulated background events passing the SR selection requirements for a given analysis channel predominantly originate from \ttjets production, with contributions of less than 7\% from \QCDjets processes, and a few percent from the remaining minor backgrounds.

\section{The ABCDisCoTEC method}
\label{sec:discoTEC}

The ABCD method~\cite{D0:1994wmk,DataAnaHEP} is a common background estimation technique that uses data CRs to extrapolate the background into the signal region. The method's success depends on choosing two quantities that are independent from each other and can distinguish between signal and background.
Having chosen two independent quantities, four regions are constructed in a two-dimensional (2D)
plane with each quantity defining an axis of the plane.
These regions are created by imposing selection criteria on both quantities. 
The CRs are subsequently referred to as \Breg, \Creg, \Dreg and
cover the domain of the plane that is mostly populated by background events.
\Breg and \Creg contain events which pass only one of the selection criteria while \Dreg contains events passing neither. 
The region subsequently referred to as \Areg is the remaining portion of the
plane where signal events are predominantly expected to reside.
All events contained in \Areg pass the selection criteria imposed on both quantities. 
Assuming independence of the two quantities defining this 2D ABCD plane,
the number of background events in the SR \Areg can be
predicted by using the \Breg, \Creg, \Dreg CRs via the relationship
\begin{equation}
    \label{eq:prediction}
    \NApred = (\NBreg\NCreg)/\NDreg,
\end{equation}
where $N_{i}$ (with $i=\Breg, \Creg, \Dreg$) is the number of
events in CR $i$, and $\NApred$ is the number of predicted background
events in the SR \Areg.

In practice, it is not always straightforward to identify two statistically
independent quantities that also provide sufficient signal sensitivity.
If the final states of signal and background events are naturally similar
to one another (as is the case for the signal models considered in this analysis), 
then simple object- or event-level quantities may not
be sufficient to perform the ABCD method and be sensitive to signal.
Rather than relying on physical quantities for defining the ABCD plane,
one can generate two synthetic quantities satisfying the ABCD criterion above using
the so-called Double Disco NN~\cite{ABCDmethod_2021}, which includes 
a statistical measure of nonlinear dependence known as distance
correlation (DisCo)~\cite{DisCo_2014} in the loss function to impose independence of its two simultaneously trained binary classifiers.
The \discotec method used in this paper augments the Double DisCo NN
with additional features including a novel loss term that directly minimizes the
nonclosure of the ABCD method using a metric 
proportional to $(\NAreg\NDreg - \NBreg\NCreg)^2$, where $\NAreg$ is the number of events in the SR \Areg. 
A more detailed description of the \discotec NN model can be found in Ref.~\cite{CMS:2025cvw}.

\section{Background estimation}
\label{sec:bkg} 

\subsection{Implementing the \discotec method for the \texorpdfstring{\ttbar}{ttbar} background}
\label{sec:discostealth}

For this search, the \discotec NN model is specifically trained to
discriminate between the main \ttjets background
and RPV or stealth \syy signal models. A separate training of the
model is performed for each of the three lepton
channels and each of the signal models, resulting in a total of
six independent NN trainings. For each training,
a mixture of simulated signal events with \mstop ranging
from 300 to 1400\GeV in 50\GeV increments is used.
To make the trained NNs more robust to potential differences between simulation and data,
the training samples include not only events from the nominal simulation, but also events
simulated with systematic variations in the physics modeling and event reconstruction.
Simulated \ttjets events include variations in the matching scale between the matrix
element and parton shower (ME-PS)~\cite{Odaka_2015}, the UE tune~\cite{Sirunyan_2020},
and the modeling of color reconnection in the parton shower in \PYTHIA~\cite{TuneCP5}, all of which have an impact on \njets. 
Variations in jet energy resolution (JER) and scale (JES) are included for both
simulated \ttjets and signal events.
Both up- and down-type variations (i.e. ${\pm} 1$ standard deviation for each systematic uncertainty) are included in the training set.
Finally, to test the NN model's performance and generalization, 80\% of the available simulated events are used for training, while the remaining 20\% are reserved for validation.

All NN input variables are computed in an approximate center-of-mass frame defined by all
jets in the event with $\pt > 30\GeV$ and $\abs{\eta} < 5$. Properties for the
first six (\twol channel), seven (\onel channel), or eight (\zerol channel) jets, including their four-vectors
and flavor tagging information, are used.
Likewise, for the \onel and \twol channels, the four-vector(s) of the lepton(s) are used as inputs.
In addition to these low-level quantities, higher-level quantities are input to the NN,
including Fox--Wolfram moments~\cite{FoxWolfram:1978}, the eigenvalues of the jet
energy-momentum tensor~\cite{JetMTensor:1970}, and the two four-vectors of reconstructed
event ``hemispheres,'' which are constructed through the object clustering method
described in Refs.~\cite{Chatrchyan_2012,PTDR2}.  
For background (signal), each clustered hemisphere typically contains the
decay products of one of the two top quarks (squarks) in the event. 

As mentioned previously, \njets is the main feature distinguishing signal from background,
and sensitivity to signal is maximized at high \njets.
To ensure that the training of the NN model is not dominated by the large number
of low-\njets events in the simulation, all \njets categories are
equally sampled during the NN training.

\begin{figure}[pth]
  \centering
  \includegraphics[width=0.47\textwidth]{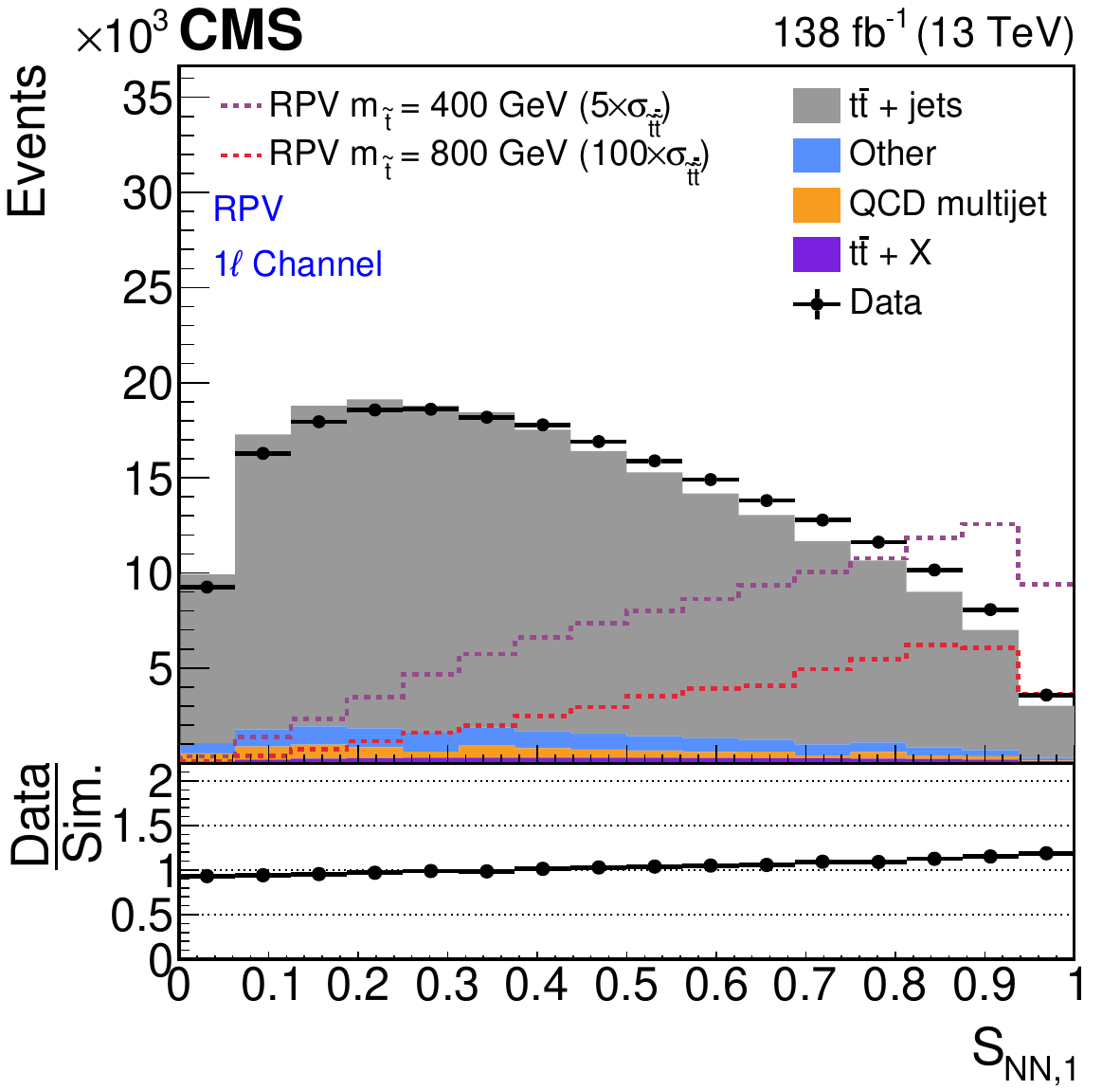}
  \includegraphics[width=0.47\textwidth]{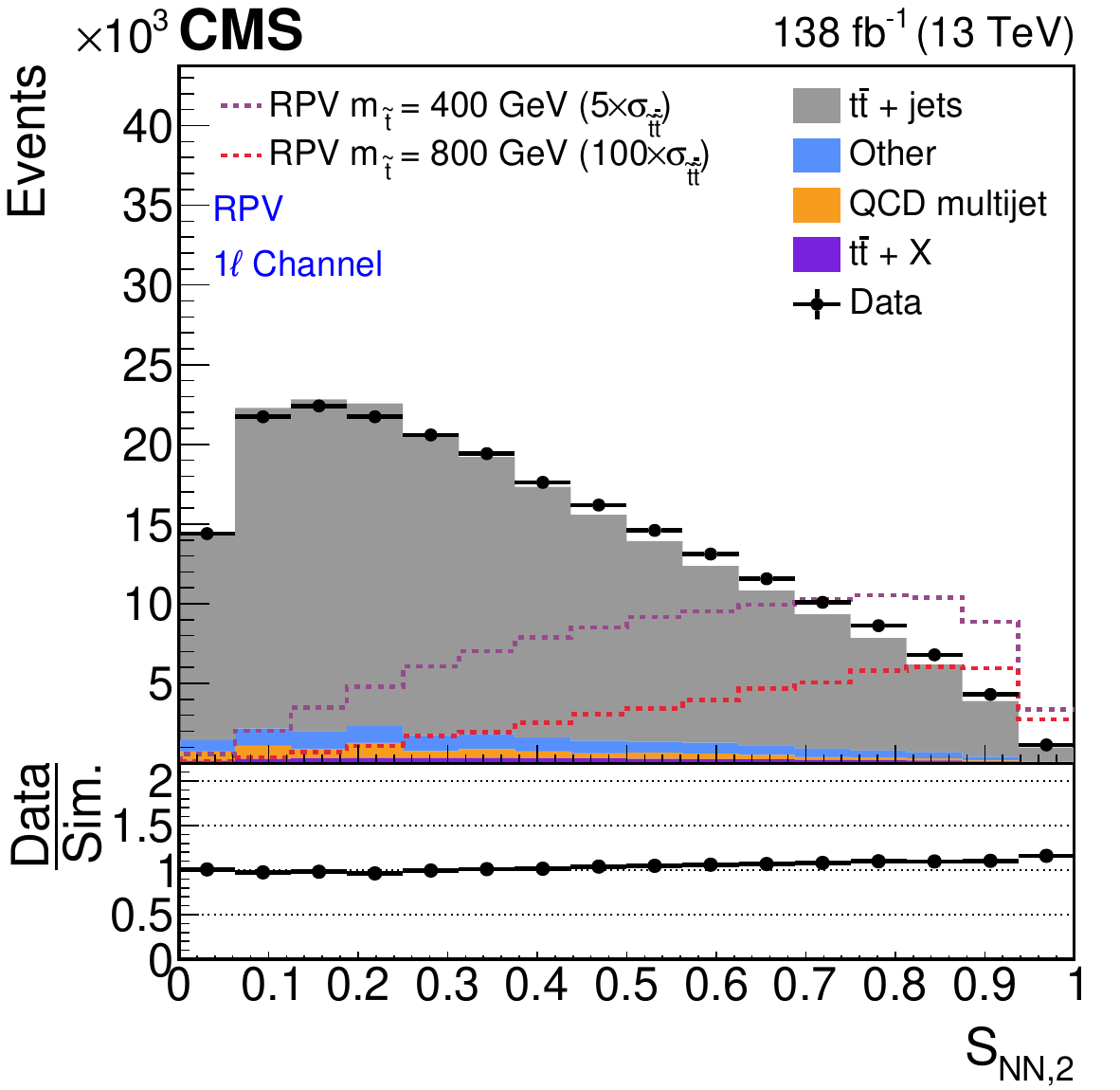}

  \includegraphics[width=0.65\textwidth]{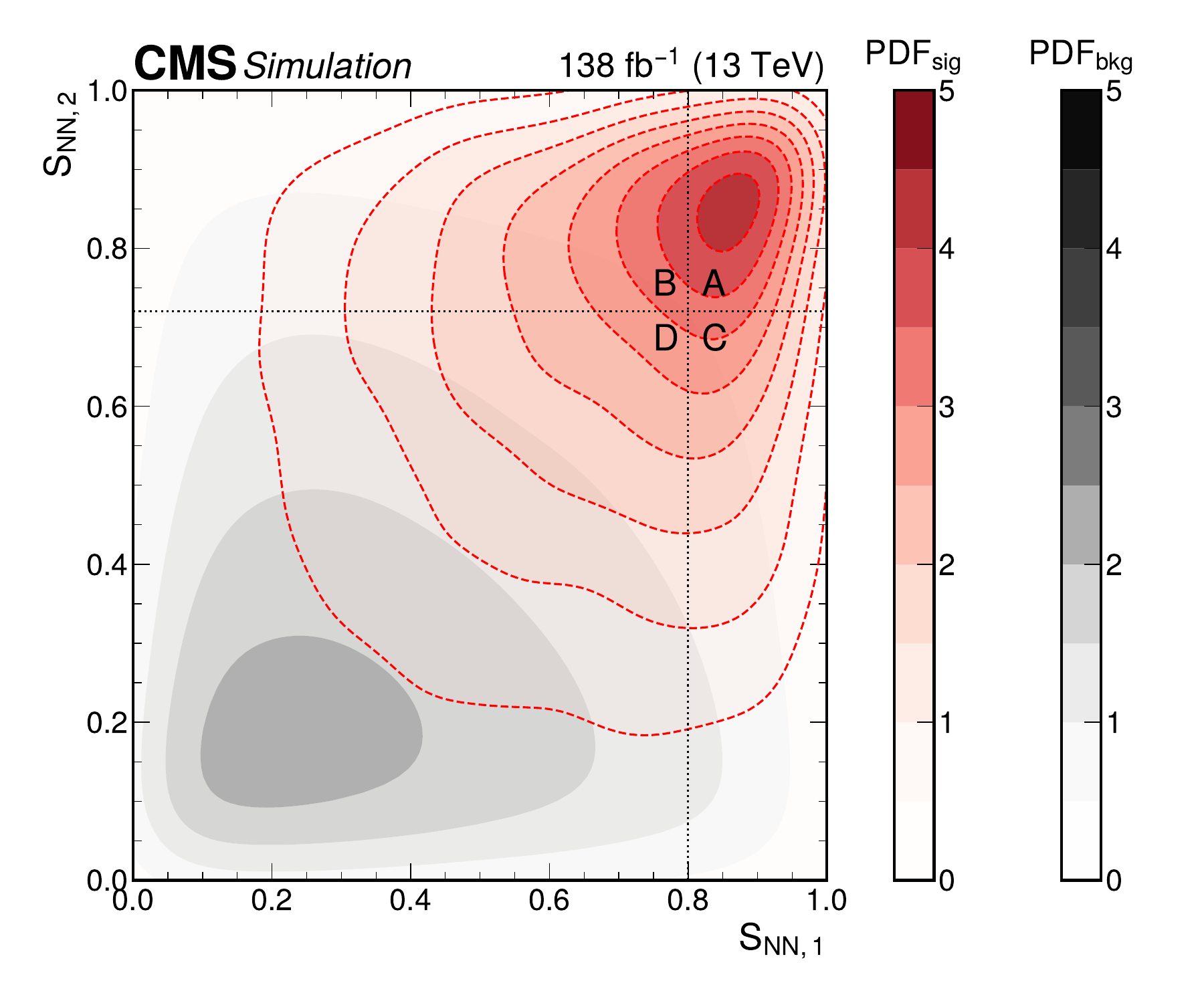}
  \caption{For the RPV signal model and all \njets categories of the \onel channel, distributions of
    \snni{1} (upper left) and \snni{2} (upper right) for data (black solid line),
    pre-fit simulated SM backgrounds (stacked filled histograms),
    and two RPV signal models with $\mstop=400$ and 800\GeV (dashed lines) are shown.
    The lower panel shows the two-dimensional probability density function distribution
    of \snni{1} and \snni{2} for simulated \ttjets events (solid gray)
    and the RPV signal model with $\mstop=800\GeV$ (dashed red).
    The ABCD bin boundaries for this
    signal model in the \onel channel
    are shown with dashed vertical and horizontal lines.
    \label{fig:NN}}
\end{figure}

\Figure{NN} shows the distributions of \snni{1} and \snni{2}, integrated over all \njets categories, 
for the training for the \onel channel and the RPV signal model for both data and simulation.
Note that as the \ttjets estimate ultimately used in the analysis
comes from the fit to data, the level of agreement between data and pre-fit simulation in the upper figures does not impact the background prediction. 
The 2D probability density functions for \ttjets and signal
simulations are smoothed with Gaussian kernel density estimation.

\subsection{The ABCD region definition and correction factor derivation}
\label{sec:optimization}

The optimal values of \snni{1} and \snni{2} for dividing
the \snni{1}-\snni{2} plane into \Areg, \Breg, \Creg, \Dreg
regions (ABCD bin boundaries) are determined with a
grid search over \snni{1} and \snni{2} considering four
figures of merit (FOMs): the expected signal significance and the upper
limit on the signal cross section for signal models with \mstop
of 400 and 800\GeV.
For each point in the scans, the complete
simultaneous fit with full treatment of
systematic uncertainties is performed in simulation.
For each lepton channel and signal model, we use two bin boundary choices
optimized separately for low and high \mstop in order to improve the performance across the full considered \mstop range;
the low-mass boundary
is used for RPV (stealth \syy) signals with masses less than 625 (675)\GeV. 
The FOM used for determining the optimal low- (high-)mass 
boundaries is the signal significance (cross section limit) for the model
with \mstop of 400 (800)\GeV.
The significance is chosen as the FOM for low \mstop to target the region in which 
the previous analysis observed an excess for the RPV model with $\mstop=400\GeV$ described in \Sec{intro}.
Finally, for each lepton channel and signal model, the same bin
boundary is used for all \njets categories. A summary of the
optimal bin boundaries is shown in \Tab{binedges}.

\begin{table}[tp]
  \centering
  \topcaption{Values of  $(\snni{1},\snni{2})$ defining the ABCD bin
    boundaries for the low- and high-mass optimizations for each
    lepton channel and signal model.}
  \begin{tabular}{l l l l}
    Signal model & Lepton channel & Low-mass boundaries & High-mass boundaries \\ \hline
    RPV & \zerol & (0.52, 0.54) & (0.74, 0.80) \\
    RPV & \onel & (0.84, 0.42) & (0.80, 0.72) \\
    RPV & \twol & (0.52, 0.58) & (0.50, 0.50) \\
    \syy & \zerol & (0.76, 0.70) & (0.54, 0.56) \\
    \syy & \onel & (0.44, 0.42) & (0.68, 0.82) \\
    \syy & \twol & (0.40, 0.42) & (0.48, 0.48) \\
  \end{tabular}
\label{tab:binedges}
\end{table}

To account for residual correlation between \snni{1} and \snni{2}, a correction factor $\kappa$
is obtained from \ttjets simulation and is applied to the \ttjets prediction
in the maximum likelihood fits. 
A separate $\kappa$ factor is calculated from the ratio of observed and predicted
\Areg region events for each \njets category, signal model, lepton channel,
and low- or high-mass range:
\begin{equation}
  \kappa = \frac{N_{\Areg,\,\text{MC}}}{N_{\Areg,\,\text{pred.}}} = \frac{N_{\Areg,\,\text{MC}}\,
    N_{\Dreg,\,\text{MC}}}{N_{\Breg,\,\text{MC}}\,N_{\Creg,\,\text{MC}}}, 
  \label{eq:closureCorrection}
\end{equation}
where $N_{\Areg,\,\text{pred.}}$ is obtained from \Equ{prediction}.
For a perfect \discotec training, each $\kappa$ would be
equal to one. Values for all sixty $\kappa$
factors range from 0.875 to 1.887 with statistical
uncertainties ranging from 0.008 to 0.12.

\subsection{Maximum likelihood fit}
\label{sec:thefit}

A set of simultaneous binned maximum-likelihood fits (over the range of \mstop)
is performed for each signal model and lepton channel for a total of six sets of fits.
In addition, a combined fit over all three lepton channels is performed for each signal model.
Each fit estimates the signal and background yields in twenty analysis
regions comprising the four \Areg, \Breg, \Creg, \Dreg regions in each of five \njets categories.
The number of jets in the \njets categories for each lepton channel range from
8 to ${\geq}12$ jets for the \zerol channel,
7 to ${\geq}11$ jets for the \onel channel,
and 6 to ${\geq}10$ jets for the \twol channel.
In each fit, the \ttjets background yields for the twenty analysis regions are
included as floating parameters $b^{j}_i$ with $j$ denoting \njets category
and $i$ denoting ABCD region. The four $b^{j}_i$ yields in each \njets category
are subject to the ABCD constraint from \Equ{prediction} including
the correction factor $\kappa^{j}$ from \Equ{closureCorrection} as
\begin{equation}
    \label{eq:tt_fit_component}
    b^{j}_\Areg = \kappa^{j} (b^{j}_\Breg b^{j}_\Creg)/b^{j}_\Dreg.
\end{equation}
Each fit includes a single free parameter determining the overall
signal normalization (referred to as the signal strength) with the
distribution of signal events across the twenty analysis
regions fixed according to simulation.
In addition to the $b^{j}_i$ parameters and the signal strength, 
each fit includes constrained nuisance parameters which are allowed to vary, accounting for a range of
systematic uncertainties.

For each fit, the \QCDjets background in each of the twenty
analysis regions is fixed according to an estimate from
the corresponding signal-depleted \QCDjets data CR and using the corresponding ABCD bin boundaries for each channel, signal, and mass optimization.
A transfer factor relating the number of simulated \QCDjets
events in the SRs and CRs is used to scale
the \QCDjets component of data in the CR 
for use as the analysis region prediction.
This transfer factor ranges from 0.0002 to 0.4 over all analysis regions and fits.
The remaining minor backgrounds are fixed according to the simulation.

\section{Systematic uncertainties}
\label{sec:syst}

The key assumption governing the \ttjets background estimation is that the ABCD relationship holds.
By definition, the correction factor $\kappa$ defined in \Equ{closureCorrection} ensures that this is the case in simulation.
We evaluate the residual nonclosure of the ABCD relationship in data (after application of the $\kappa$ correction)
and determine a related systematic uncertainty with a two-fold process performed separately for
each lepton channel (\zerol, \onel, \twol), signal model (RPV or stealth \syy),
and ABCD boundary choice (low- or high-mass).

\begin{figure}[pth]
  \centering
  \includegraphics[width=0.8\textwidth]{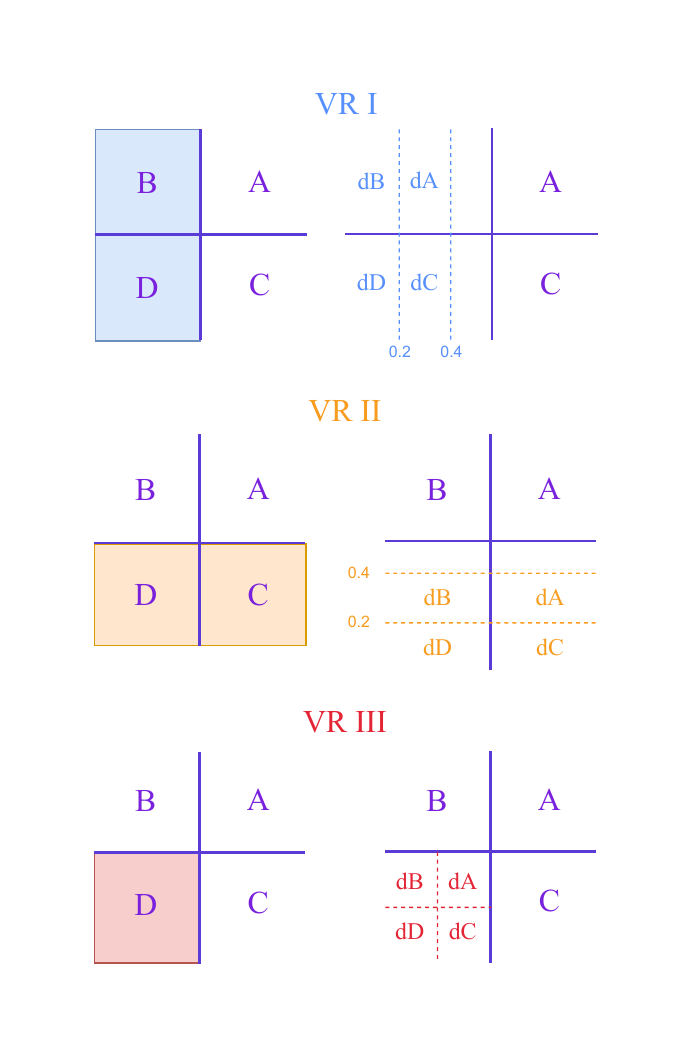}
    \caption{Illustrations of the three validation regions that are created by partitioning the \snni{1}-\snni{2} plane.
       VR\,I (VR\,II), shown in the upper (middle) row, is a division of the \Breg and \Dreg (\Creg and \Dreg) regions.
       VR\,III (lower row) is a division of the \Dreg region.
       Each VR is divided into four subregions (d\Areg, d\Breg, d\Creg, d\Dreg) that are used to perform the validation
       of the nonclosure of the \snni{1}-\snni{2} plane. 
       The subregion boundaries shown in the figure are the starting points of the stepping procedure explained in the text. 
    \label{fig:valRegions}}
\end{figure}

First, we use three signal-depleted data validation regions (VRs) in which the ABCD relationship should hold to 
evaluate a residual nonclosure metric \nonclosure, defined as
\begin{equation}
  \nonclosure = 1-\frac{\kappa \NApred}{\NAreg} = 1-\frac{\kappa \NBreg\NCreg}{\NAreg\NDreg}. 
  \label{eq:nonclosure}
\end{equation}
As illustrated in Fig.~\ref{fig:valRegions}, the region VR\,I (VR\,II) is defined as approximately the combination of the nominal $\Breg\Dreg$ ($\Creg\Dreg$) region,
and thus spans the full \snni{2} (\snni{1}) range.  The ABCD subregions of VR\,I (VR\,II) are defined by
bisecting VR\,I (VR\,II) on the \snni{1} (\snni{2}) axis while keeping the nominal
boundary on \snni{2} (\snni{1}) constant. The region VR\,III comprises the \Dreg region and
is bisected on both \snn axes to create the ABCD subregions.
In order to address potential dependence of \nonclosure on the location of the ABCD bin boundaries,
the boundaries of each VR (and its ABCD subregions) are moved in steps towards the nominal ABCD bin boundaries  
and \nonclosure is calculated at each step as described in Ref.~\cite{CMS:2025cvw}.
The $N_i$ counts used in each calculation are taken from the data after
subtraction of non-\ttbar background according to simulation.
Only those ABCD subregions with an expected signal fraction
less than 5\% are considered in the nonclosure study. 
The systematic uncertainty associated with the residual nonclosure in data (data-based nonclosure systematic uncertainty) for a
given search channel, signal model, and ABCD bin boundary
is taken as the largest deviation of \nonclosure from unity from all
three VR scans for the lowest \njets category.
Categories at higher \njets do not have a sufficient number of steps in the procedure passing the 5\% signal fraction requirement to provide an uncertainty estimate. 
Therefore, the same uncertainty is used for all \njets categories.
Values of this uncertainty range from 3 to 15\%.

Second, to account for potential additional effects on the residual nonclosure in data
at high \njets, we use simulated \ttjets samples in which various aspects of \ttjets
modeling are varied including ISR and FSR, ME-PS matching, UE tune, color reconnection,
JES, and JER. The same process for calculating $\kappa$ is applied to these
varied \ttjets samples and any difference in $\kappa$ between the nominal and
varied samples is taken as a systematic uncertainty for cases
in which the change in $\kappa$ exceeds the statistical
uncertainty summed in quadrature with the residual nonclosure
systematic uncertainty described above;
the only modeling variation for which $\kappa$ meets this requirement is
that related to FSR.
This uncertainty has a typical size of 10--20\% on the $\ttjets$ normalization, but can be as
large as 96\% for the highest \njets category in the \zerol region.
The statistical uncertainty in the nominal $\kappa$ factor is also
included as a systematic uncertainty in the \ttjets prediction.

\begin{figure}[tbp]
  \centering
    \includegraphics[width=\textwidth]{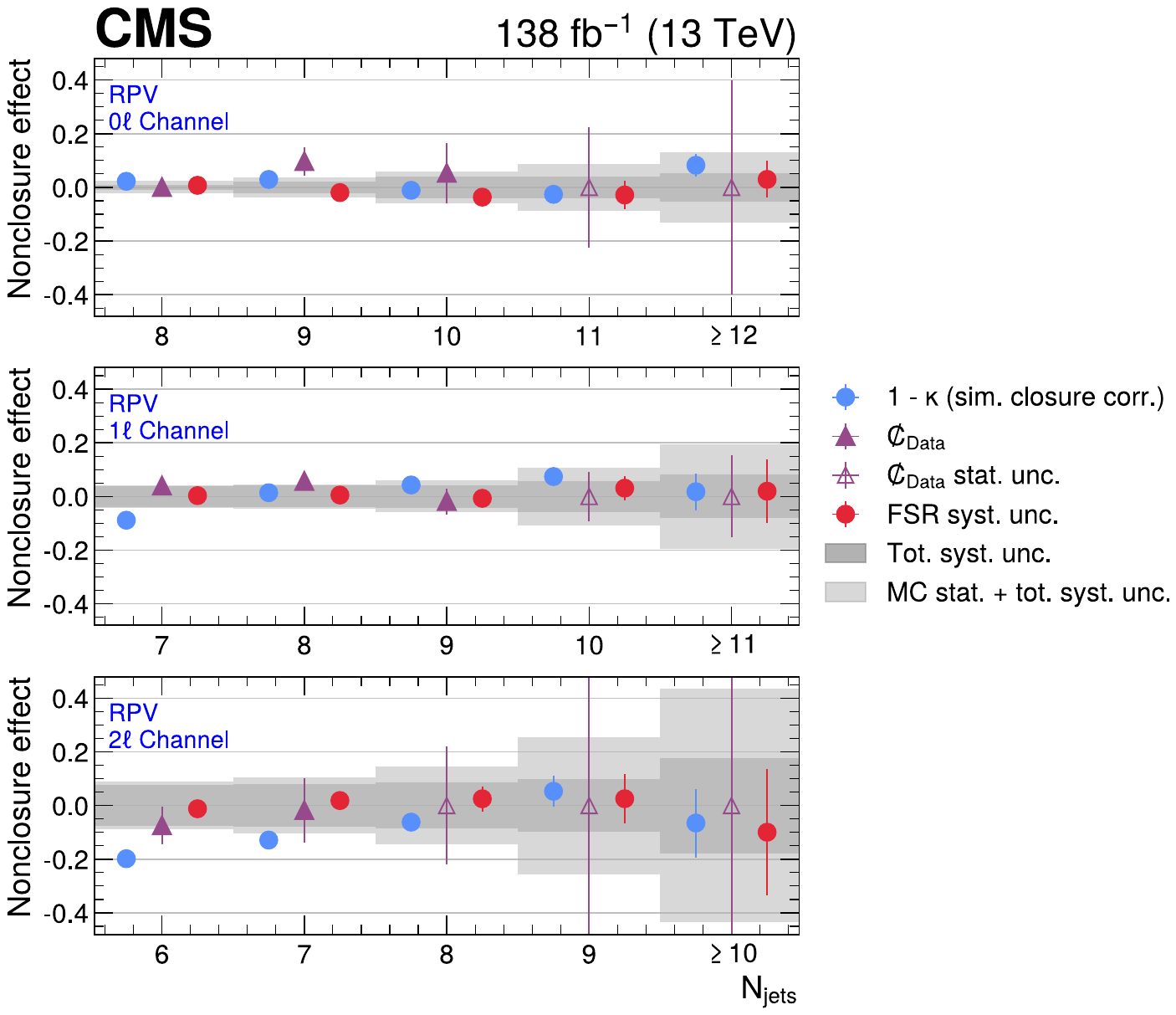}
  \caption{One minus the simulation-based closure correction $\kappa$ (blue circle),
    the residual nonclosure \nonclosure in data (solid purple triangle),
    and the FSR systematic uncertainty (red circle) are shown for each \njets category
    for the \zerol (upper), \onel (middle), and \twol (lower panel) channels.
    All values correspond to the low-mass optimization for the RPV signal model.
    The value of \nonclosure shown is the maximum value of the stepping procedure described in the text. 
    Note the data-based nonclosure systematic uncertainty is defined to be the nonclosure in the lowest \njets bin
    for each of the channels. 
    Open purple triangles show the statistical uncertainty in \nonclosure in data for categories
    in which all VR and ABCD boundaries have a signal fraction exceeding 5\%. 
    All data-based nonclosure values are calculated after applying the
    simulation-based closure correction, such that an observed nonclosure
    of zero signifies identical modeling of the \snni{1}-\snni{2} correlation in simulation and data.
  \label{fig:closureComparison}}
\end{figure}

Figure~\ref{fig:closureComparison} shows
the maximum \nonclosure in data for each \njets category across the VRs and ABCD bin boundaries with
expected signal fraction less than 5\% for the RPV low-mass optimization.
For \njets categories in which no VR and ABCD bin boundaries satisfy this requirement, the
statistical uncertainty in \nonclosure for the boundaries listed in \Tab{binedges} is shown. 
Also shown are the deviations from one of the simulation-based closure correction $\kappa$, the FSR systematic uncertainty,
and the total statistical and systematic uncertainty for the \ttjets prediction. 
In all cases, \nonclosure
is comparable to the total systematic uncertainty.
In addition, we observe that \nonclosure is small even if $\kappa$ is larger than the typical
size of the uncertainties, as seen in the figure for the \twol channel, indicating that the simulation
models the data well for any residual \snni{1}-\snni{2} correlations. 
Across all channels and optimizations, the largest $\kappa$ factor has a value of 1.887 and is found for the RPV \zerol channel high-mass optimization, 
resulting in a value of only 8\% for the data-based nonclosure systematic uncertainty.

For minor background processes and signal, systematic uncertainties are estimated
to account for both physical mismodeling of events and detector effects.
These include uncertainties in JES and JER~\cite{CMS:2011shu}, PDFs, \PQb and \PQt~tagging efficiencies and misidentification rates, lepton identification, trigger efficiencies, pileup, renormalization and factorization scales ($\mu_\mathrm{R}$, $\mu_\mathrm{F}$), the integrated luminosity~\cite{CMS:LUM-17-003, CMS:LUM-17-004, CMS:LUM-18-002}, and in the correction factor for the ECAL timing shift~\cite{CMS:2020cmk}, referred to as trigger prefiring.
These uncertainties are applied to each of the analysis regions to allow for adjustments in the normalization and shape of each of these processes. 
All systematic uncertainties originating from the same source are correlated across all \njets and ABCD bins, search channels, and processes.
Minor background processes normalizations are assigned an additional 20\% uncertainty to account for discrepancies between LO and NLO cross section calculations~\cite{Kim:2024ppt}.

\Table{systematics} shows the relative size of each systematic uncertainty in the expected event yield for the background and signal processes across the three search channels for the RPV SUSY model and the ABCD boundaries optimized for low \mstop. 
The ranges shown for each systematic uncertainty account for the difference in their value across $\njets$ categories and ABCD regions. 

\begin{table}[tph]
  \centering
  \topcaption{Magnitude of systematic uncertainties for the \zerol, \onel, and \twol channels based on the RPV-trained NN and ABCD bin boundaries optimized for low \mstop. Reported values are in units of \% and ranges correspond to the 16th and 84th percentile for the value of a systematic uncertainty across all applicable analysis regions (ABCD regions and \njets categories). The maximum value for a given systematic uncertainty across these regions is shown in parentheses.
    Other backgrounds include the \QCDjets and minor background processes. The RPV signal model used assumes $\mstop = 550\GeV$. The systematic uncertainties based on the RPV-trained model are representative of the values obtained for the stealth \syy-trained model.}
  \cmsTable{
  \def\arraystretch{1.2}
  \begin{tabular}{l r r r r r r r r r r r}
    Source of & & \zerol & & \hspace*{0.5cm} & & \onel & & \hspace*{0.5cm} & & \twol & \\
    uncertainty (\%)      & \ttbar     & Other      & RPV signal   &  & \ttbar     & Other      & RPV signal    &  & \ttbar     & Other      & RPV signal    \\
   \hline 
    PDF                                         & \NA     & $<$1 (1)    & $<$1        &  & \NA      & $<$1 (1)    & $<$1       &  & \NA        & $<$1 (2)     & $<$1     \\
   ($\mu_\mathrm{R}$, $\mu_\mathrm{F}$) scales & \NA      & 1--8 (16)   & 0--1 (2)    &  & \NA      & 0--5 (13)   & 0--1 (2)   &  & \NA        & 0--7 (19)    & 0--1 (2)   \\
   FSR                                         & 1--3 (3) & 1--14 (56)  & 1--12 (18)  &  & 0--1 (3) & 1--12 (26)  & 1--7 (15)  &  & 0--6 (9)   & 2--21 (100)  & 1--9 (16)  \\
   ISR                                         & \NA      & 0--10 (17)  & 1--4 (5)    &  & \NA      & 0--8 (15)   & 1--4 (5)   &  & \NA        & 3--11 (17)   & 0--4 (5)   \\
   Pileup                                      & \NA      & 0--2 (12)   & 0--1 (3)    &  & \NA      & 0--1 (22)   & $<$1 (3)   &  & \NA        & 0--9 (26)    & 0--1 (9)   \\
   ABCD nonclosure                             & 3        & \NA         & \NA         &  & 5        & \NA         & \NA        &  & 7          & \NA          & \NA        \\
   $\kappa$ stat. unc.                         & 1--4 (7) & \NA         & \NA         &  & 1--7 (8) & \NA         & \NA        &  & 5--15 (19) & \NA          & \NA        \\
   QCD transfer factor                         & \NA      & 1--4 (16)   & \NA         &  & \NA      & $<$1 (1)    & \NA        &  & \NA        & $<$1         & \NA        \\
   JES                                         & \NA      & 4--18 (100) & 0--10 (27)  &  & \NA      & 1--18 (100) & 1--14 (19) &  & \NA        & 4--100 (100) & 1--15 (25) \\
   JER                                         & \NA      & 0--8 (23)   & 0--2 (4)    &  & \NA      & 0--5 (35)   & 0--1 (4)   &  & \NA        & 0--20 (45)   & 0--3 (9)   \\
   b tagging                                   & \NA      & 0--1 (7)    & 1--2 (3)    &  & \NA      & 0--1 (3)    & $<$1       &  & \NA        & 0--1 (7)     & $<$1 (2)   \\
   t tagging                                   & \NA      & 26--33 (42) & 26--31 (34) &  & \NA      & \NA         & \NA        &  & \NA        & \NA          & \NA        \\
   Jet trigger                                 & \NA      & 0--1 (1)    & $<$1 (1)    &  & \NA      & \NA         & \NA        &  & \NA        & \NA          & \NA        \\
   Lepton ID                                   & \NA      & \NA         & \NA         &  & \NA      & 3--3 (4)    & 3--3 (3)   &  & \NA        & 5--6 (6)     & 5--5 (6)   \\
   Trigger prefiring                           & \NA      & 0--2 (7)    & 0--2 (3)    &  & \NA      & 0--3 (6)    & 0--3 (4)   &  & \NA        & 0--3 (11)    & 0--2 (4)   \\
   Integrated lumi.                       & \NA      & 1.6         & 1.6         &  & \NA      & 1.6         & 1.6        &  & \NA        & 1.6          & 1.6        \\
   Theor. cross section                   & \NA      & 20          & \NA         &  & \NA      & 20          & \NA        &  & \NA        & 20           & \NA        \\
  \end{tabular}                             
  }
  \label{tab:systematics}
\end{table}

\section{Results and interpretation}
\label{sec:res}

\begin{figure}[p]
  \centering
    \includegraphics[width=0.90\textwidth]{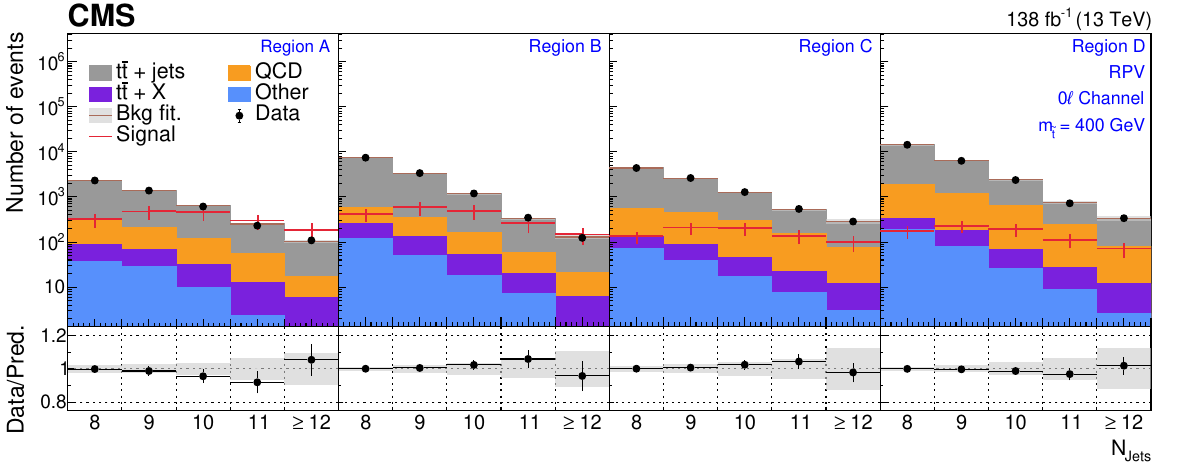}
    \includegraphics[width=0.90\textwidth]{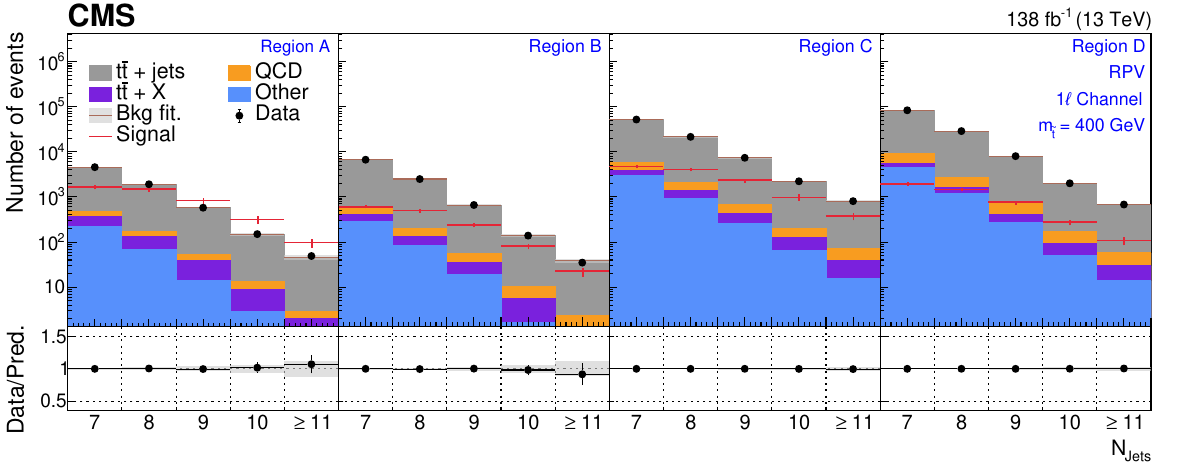}
    \includegraphics[width=0.90\textwidth]{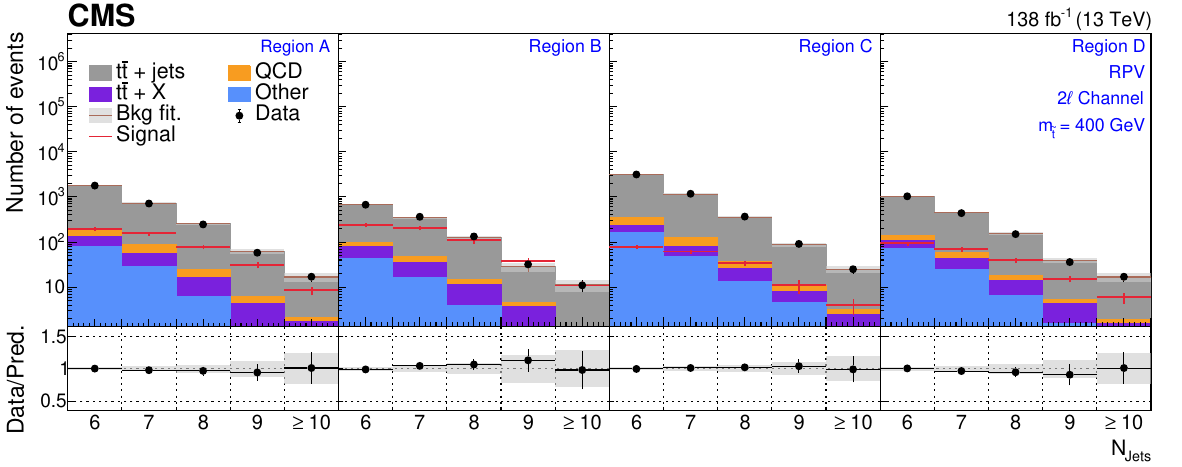}
  \caption{The \njets distributions from the background-only fits to data. Fits are run with the signal strength fixed to zero and all background process event yields are obtained from their fit predictions. The signal distribution overlaid corresponds to the RPV model with $\mstop = 400\GeV$. The graphs in the upper, middle, and lower rows correspond to the \zerol, \onel, and \twol channels, respectively. The four panels in each row show the \njets distributions for the \Areg, \Breg, \Creg, and \Dreg regions (left to right). The gray error band shows the combined statistical and systematic uncertainty from the background-only fit distributions.}
  \label{fig:backgroundOnly}
\end{figure}

The fit results for the full Run 2 data with the signal strength fixed to zero (background-only fit) are shown in Fig.~\ref{fig:backgroundOnly} for the RPV signal model with ABCD boundaries optimized for low \mstop.
The results for the stealth \syy model and the high-mass optimization are shown in Appendix~\ref{app}.  
Ratios of numbers of observed data and predicted events as a function of \njets are shown in the lower panel with error bars representing the systematic and statistical uncertainties.
Overall, good agreement is observed between the data and the prediction; no excess of events is observed. 

\begin{figure}[p]
  \centering
    \includegraphics[width=0.48\textwidth]{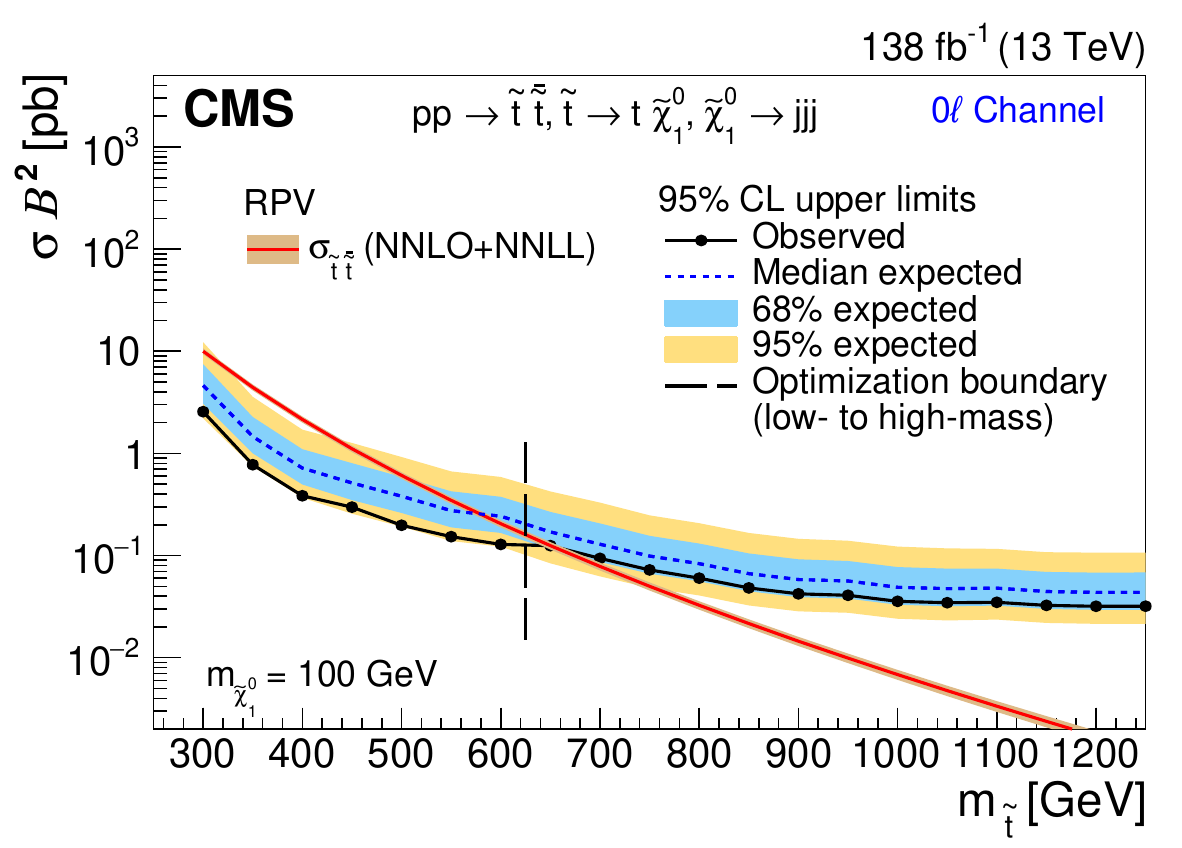}
    \includegraphics[width=0.48\textwidth]{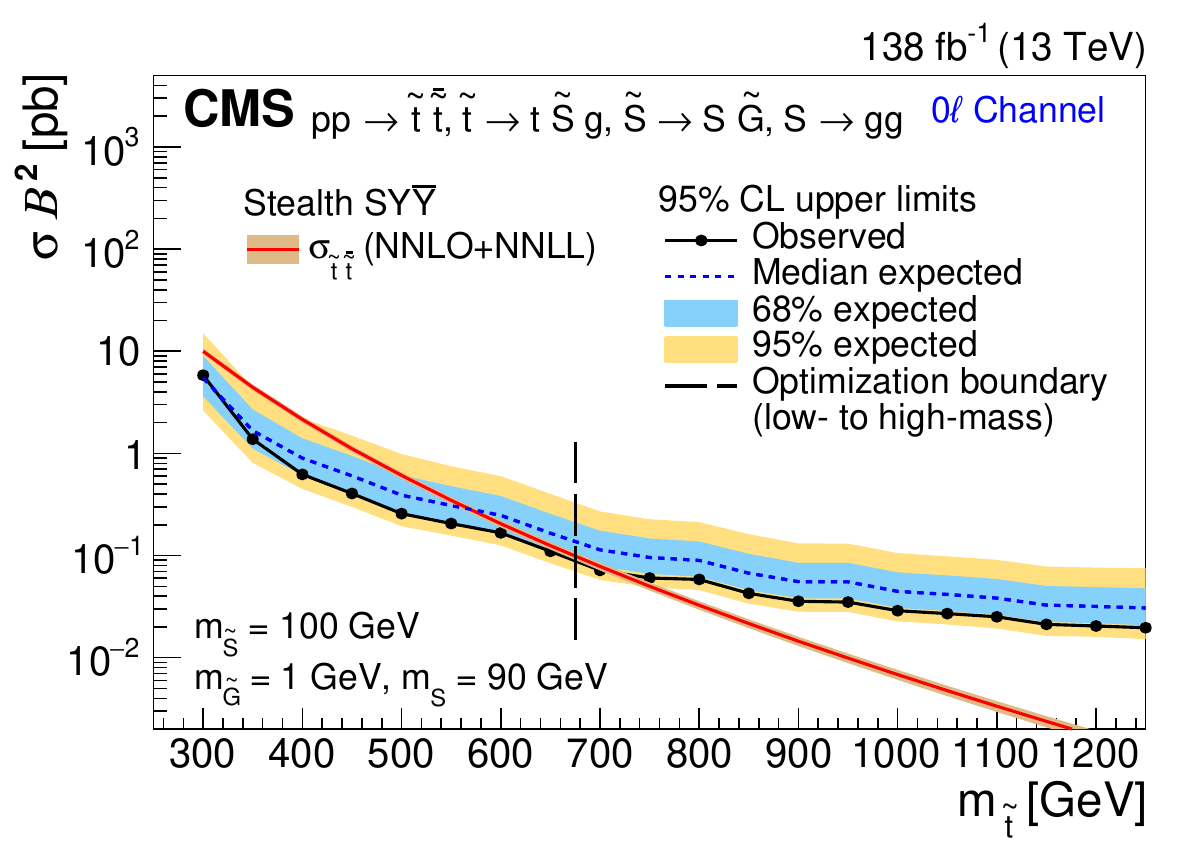}

    \includegraphics[width=0.48\textwidth]{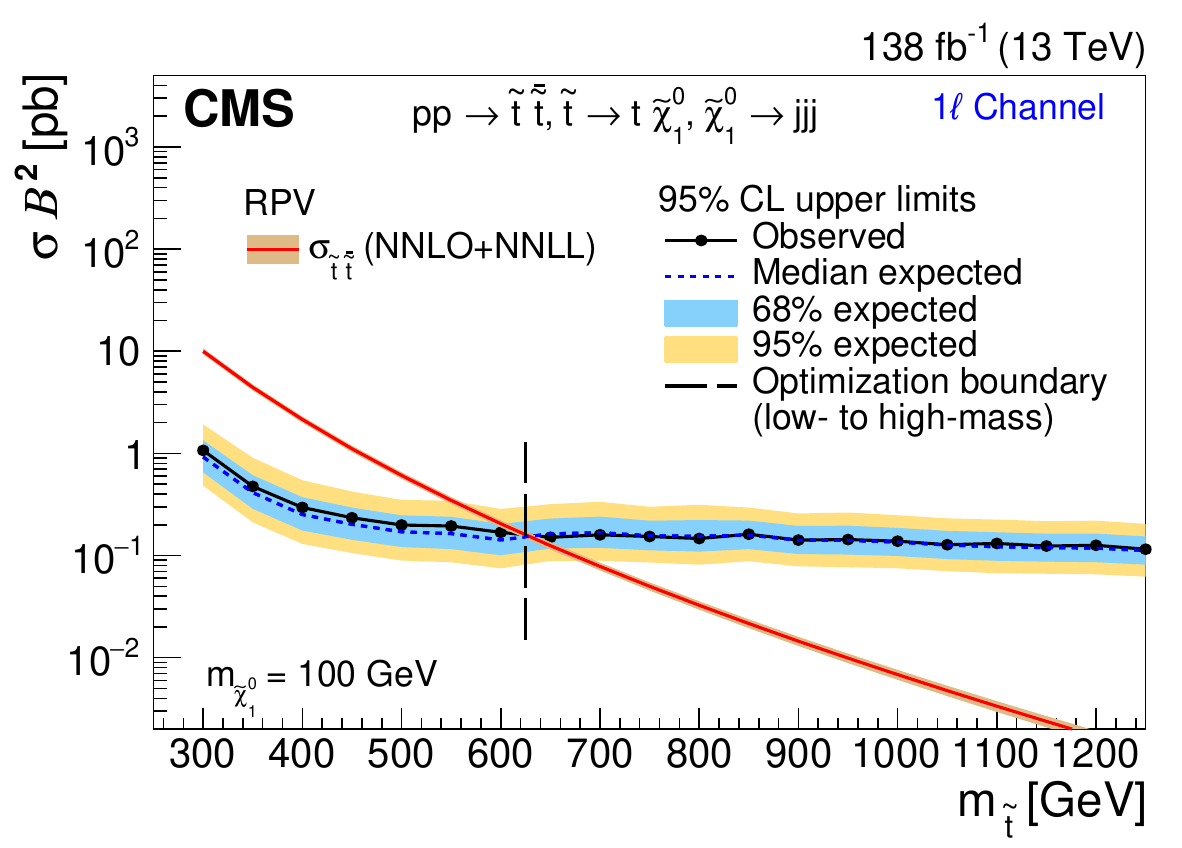}
    \includegraphics[width=0.48\textwidth]{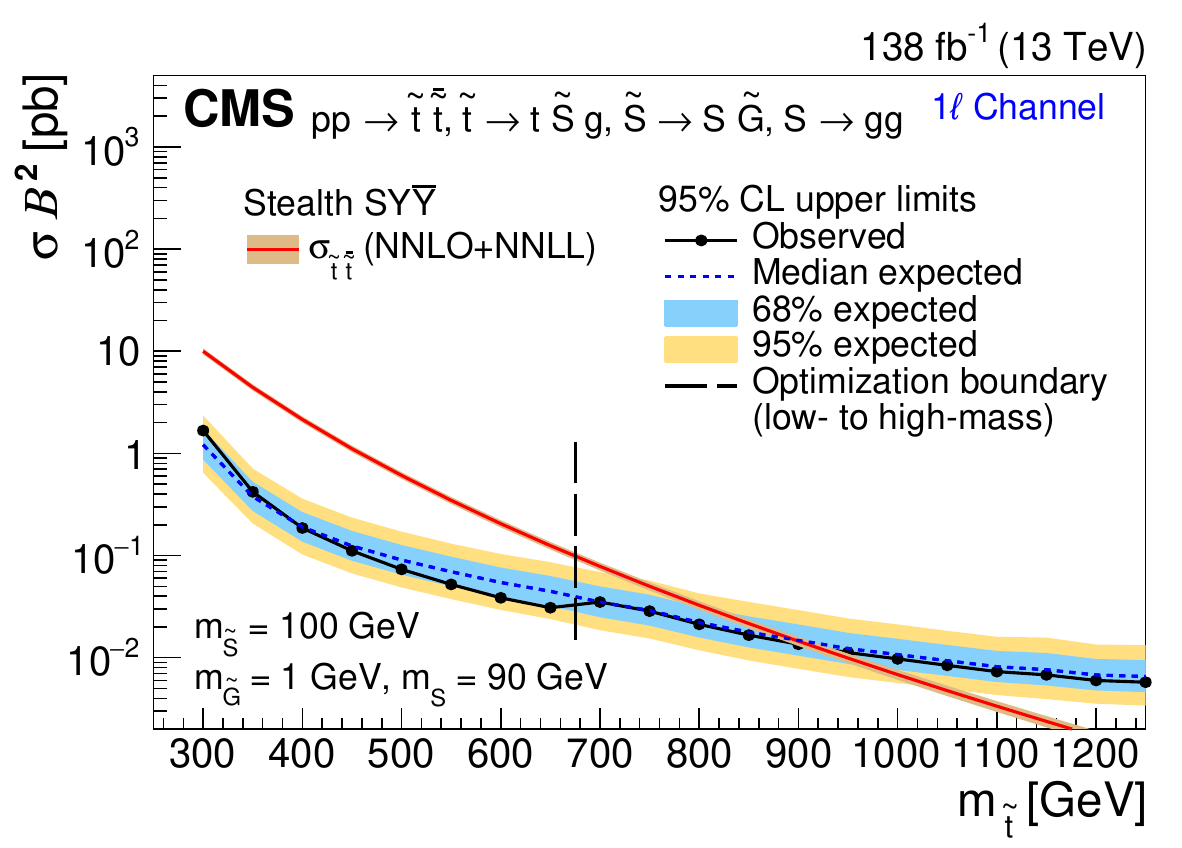}

    \includegraphics[width=0.48\textwidth]{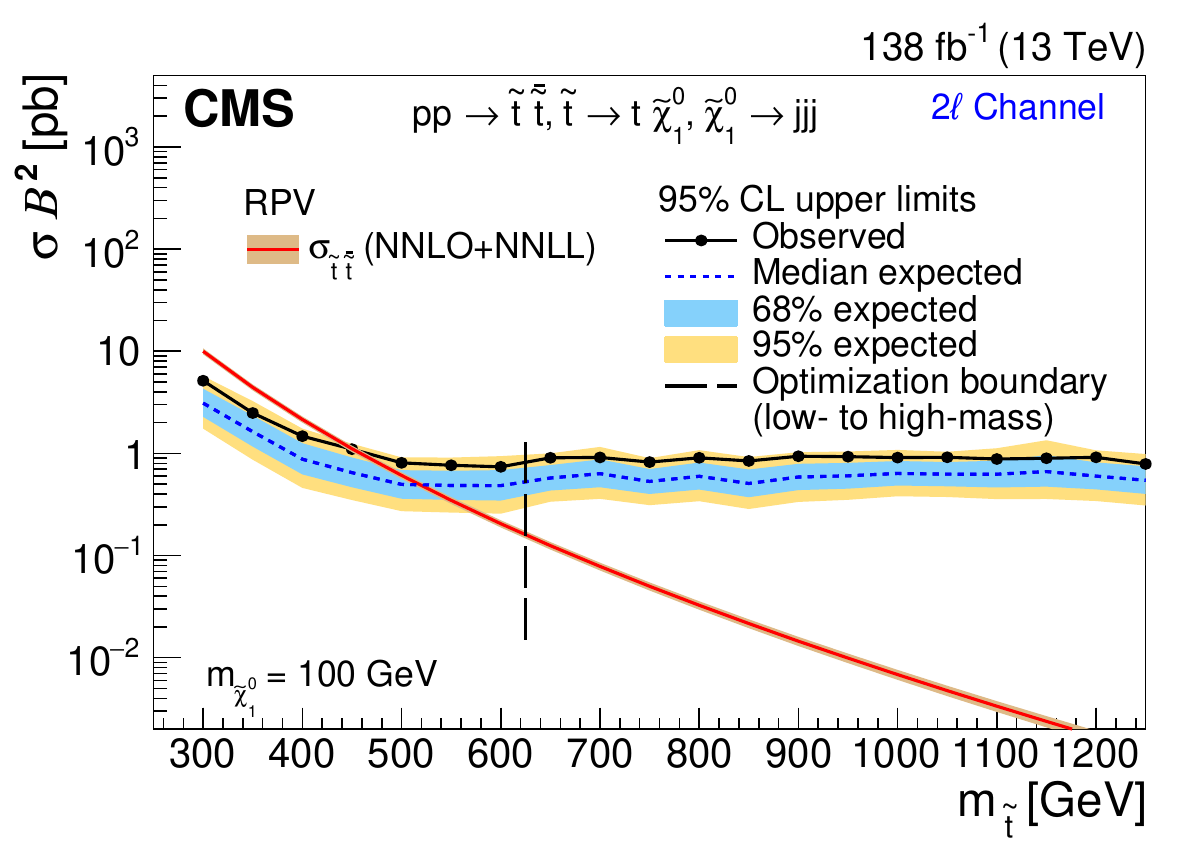}
    \includegraphics[width=0.48\textwidth]{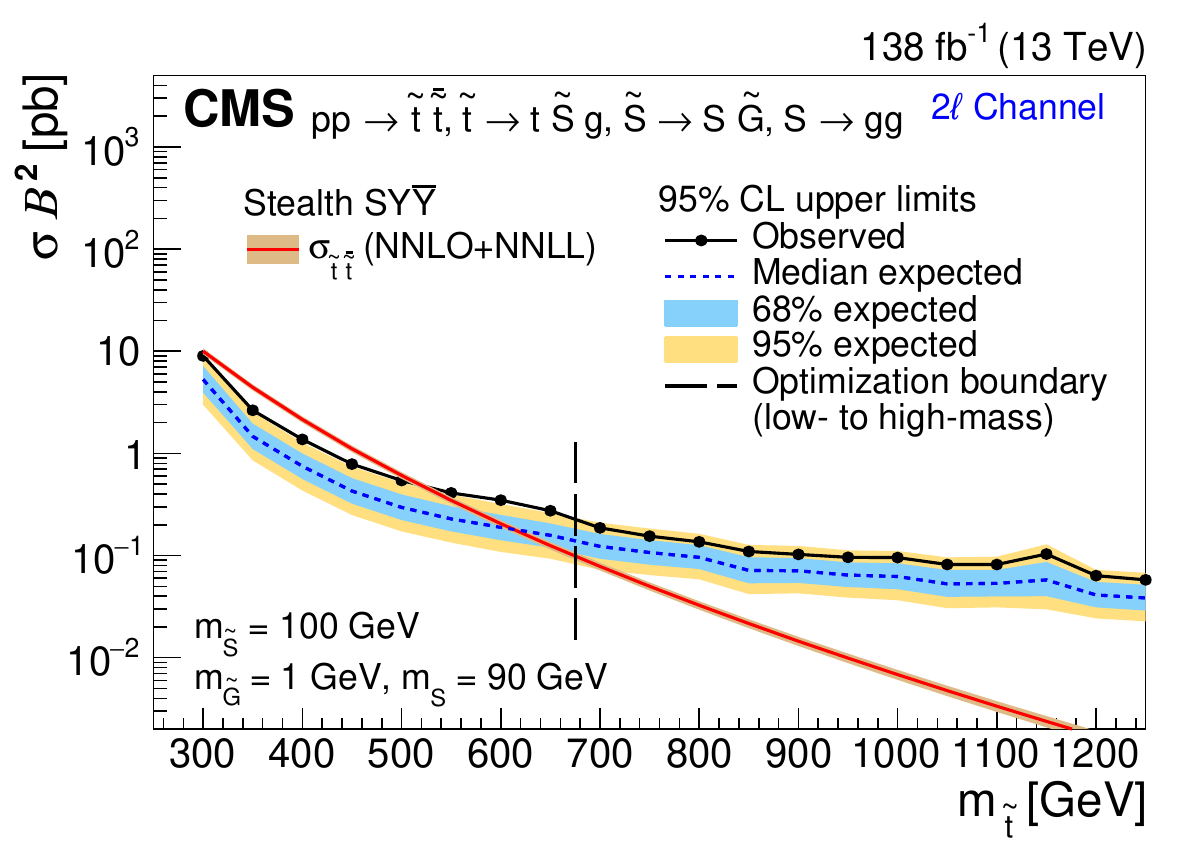}

  \caption{The 95\% \CL upper limit on \signalxsec for the RPV (left) and stealth \syy (right) SUSY signal models as a function of \mstop for the \zerol channel (upper), \onel channel (middle), and \twol channel (lower), assuming 100\% branching fraction to the considered \PSQt squark decay. The median expected limit is shown in the dashed blue line, with the 68 and 95\% intervals shown in light-blue and yellow, respectively. The observed limit is shown in black. 
  The vertical dashed line denotes the transition from the low- to high-mass optimization ABCD boundaries.
  Additionally, the theoretical cross section is shown in red with its uncertainty in light brown.}
  \label{fig:limits_perchannel}
\end{figure}

\begin{figure}[p]
  \centering
    \includegraphics[width=0.90\textwidth]{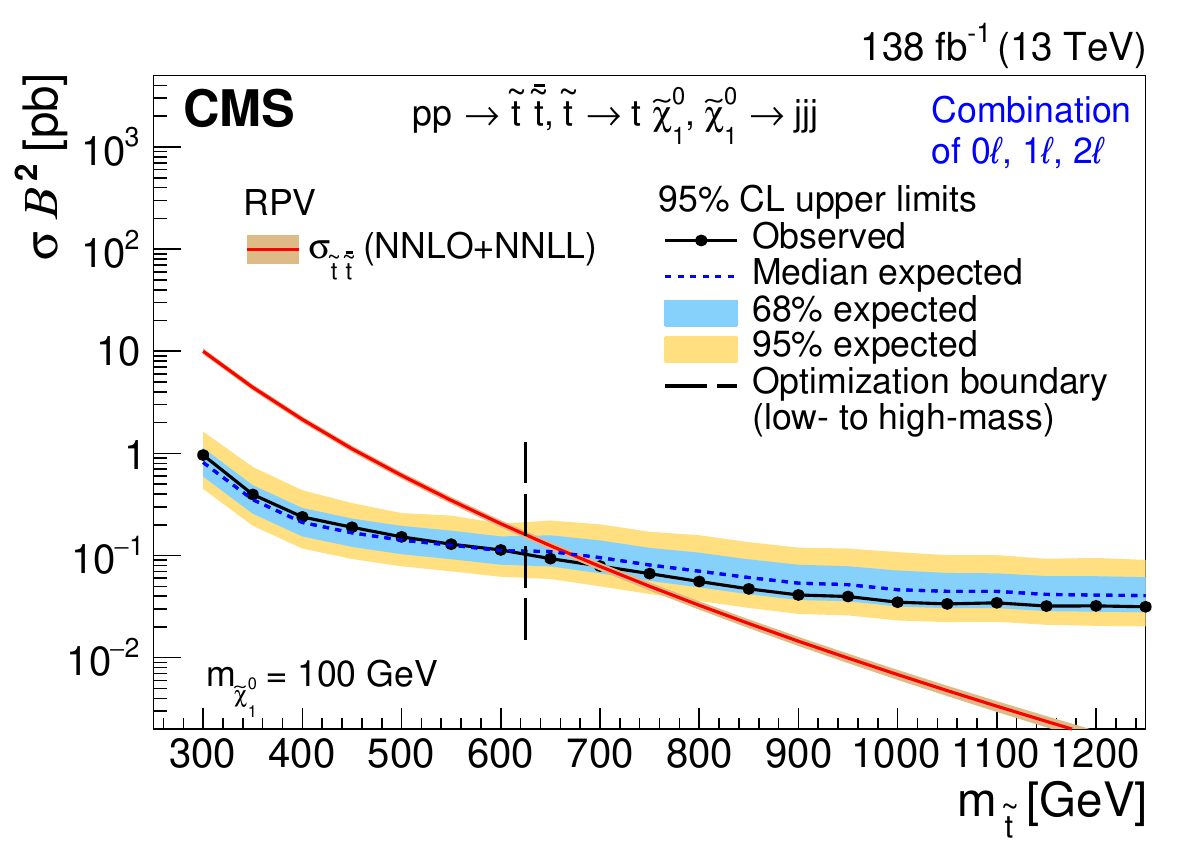}

    \includegraphics[width=0.90\textwidth]{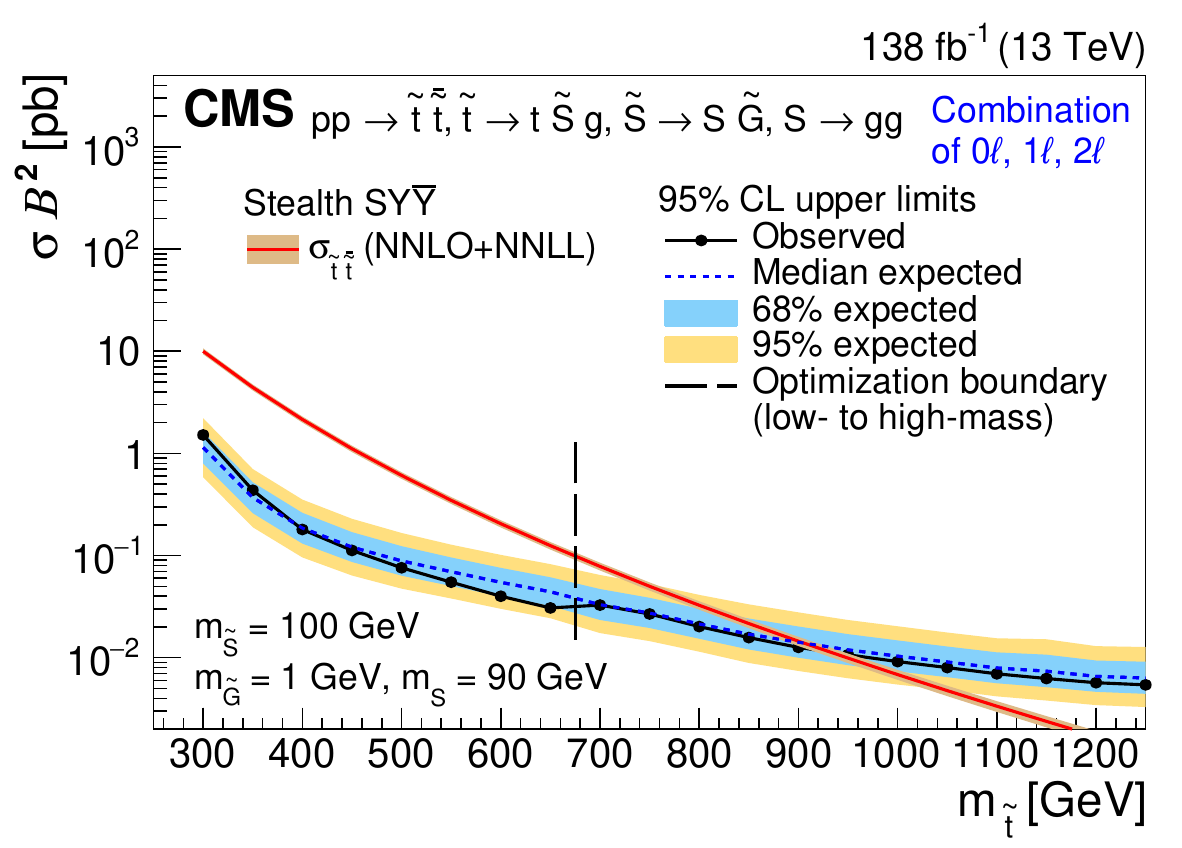}
  \caption{Observed and expected upper limits at 95\% \CL on \signalxsec for the RPV (upper) and stealth \syy (lower) SUSY signal models as functions of \mstop for the combination of all three channels, assuming 100\% branching fraction to the considered \PSQt squark decay. The median expected limit is shown in the dashed blue line, with the 68 and 95\% intervals shown in light blue and yellow, respectively. The observed limit is shown in black. The vertical dashed line denotes the transition from the low- to high-mass optimization ABCD boundaries. Additionally, the theoretical cross section is shown in red with its uncertainty in light brown.}
  \label{fig:limits_combined}
\end{figure}

The data are used to place 95\% confidence level (\CL) upper limits on the signal cross section \signalxsec  
for both the RPV and stealth \syy SUSY models according to the \CLs criterion~\cite{CLS1,CLS2}
with an LHC-style likelihood ratio~\cite{CMS-NOTE-2011-005} as the test statistic using the asymptotic approximation~\cite{Cowan:2010js}.
These results are determined using the CMS statistical analysis tool \textsc{Combine}~\cite{CMS:2024onh} and assume 100\% branching fraction to the considered \PSQt squark decay. 
The low-mass ABCD boundaries are used for the RPV (stealth \syy) signal models with $\mstop < 625\, (675)\GeV$, with the high-mass boundaries used for higher \mstop. 
Figures~\ref{fig:limits_perchannel} and \ref{fig:limits_combined} show the expected and observed upper limit on \signalxsec for the RPV and stealth \syy models with \mstop ranging from 300 to 1250\GeV for each lepton channel separately and for all three channels combined, respectively.
The theoretical cross section is also shown along with its uncertainty. 
Based on the results from the combination of all channels, \PSQt squark masses up to 700\GeV are excluded for the RPV model and up to 930\GeV for the stealth \syy model.

\section{Summary}
\label{sec:sum}

A search for the pair production of top squarks with decays to
top quarks and six additional gluons or light-flavor quarks via $R$-parity violating (RPV) or stealth \syy supersymmetric decays
is presented.
The search is performed using proton-proton collision events 
at $\sqrt{s} = 13\TeV$, corresponding to an integrated luminosity of 138\fbinv, collected by the CMS detector in 2016--2018.
Events are selected in three search channels, defined as
having at least six jets and zero, one, or two electrons or muons. 
No requirement is placed on the presence of missing transverse momentum.

This analysis is an extension of a previous search for these signatures~\cite{SUS-19-004},
which observed a deviation with a local significance of 2.8 standard deviations for a top squark mass of 400\GeV for the RPV model.
The main improvements of this search are the addition of the zero- and two-lepton channels as well as the inclusion of
a novel, neural-network-based background estimation method referred to as \discotec~\cite{CMS:2025cvw}. 
The key feature of this new method is the creation of two uncorrelated neural network variables that can be used
with an ABCD-style background estimation method. 
The backgrounds are estimated from a simultaneous binned likelihood fit to all search channels in several categories of jet multiplicity,
with the contribution from the main \ttjets background constrained via the ABCD relationship that encodes the independence between the two neural network variables. 
This new background estimation method, compared to that of Ref.~\cite{SUS-19-004}, reduces the dependence of the analysis on uncertainties related to the modeling of the jet multiplicity spectrum. 
With this alternate background estimation technique, we achieve an improvement in signal sensitivity at low top squark mass.
For example, the expected upper limit on the production cross section improves by a factor of 1.53 for a top squark mass of 400\GeV for the RPV model.
Overall, good agreement between the data and the background prediction is observed, and the deviation observed previously is not confirmed. 

The results are interpreted using two top squark decay topologies that
generate signatures with low missing transverse momentum: the RPV and stealth \syy models. 
For the RPV model with top squark decays to a top quark and neutralino,
with the subsequent decay of the neutralino to three light-flavored
quarks, top squark masses up to 700\GeV are excluded at 95\% confidence level. 
Similarly, top squark masses of up to 930\GeV are excluded in the context of the stealth \syy model where top squarks
decay to a top quark, gluons, and a soft gravitino via a hidden sector.

\begin{acknowledgments}
  We congratulate our colleagues in the CERN accelerator departments for the excellent performance of the LHC and thank the technical and administrative staffs at CERN and at other CMS institutes for their contributions to the success of the CMS effort. In addition, we gratefully acknowledge the computing centers and personnel of the Worldwide LHC Computing Grid and other centers for delivering so effectively the computing infrastructure essential to our analyses. Finally, we acknowledge the enduring support for the construction and operation of the LHC, the CMS detector, and the supporting computing infrastructure provided by the following funding agencies: SC (Armenia), BMBWF and FWF (Austria); FNRS and FWO (Belgium); CNPq, CAPES, FAPERJ, FAPERGS, and FAPESP (Brazil); MES and BNSF (Bulgaria); CERN; CAS, MoST, and NSFC (China); MINCIENCIAS (Colombia); MSES and CSF (Croatia); RIF (Cyprus); SENESCYT (Ecuador); ERC PRG, RVTT3 and MoER TK202 (Estonia); Academy of Finland, MEC, and HIP (Finland); CEA and CNRS/IN2P3 (France); SRNSF (Georgia); BMBF, DFG, and HGF (Germany); GSRI (Greece); NKFIH (Hungary); DAE and DST (India); IPM (Iran); SFI (Ireland); INFN (Italy); MSIP and NRF (Republic of Korea); MES (Latvia); LMTLT (Lithuania); MOE and UM (Malaysia); BUAP, CINVESTAV, CONACYT, LNS, SEP, and UASLP-FAI (Mexico); MOS (Montenegro); MBIE (New Zealand); PAEC (Pakistan); MES and NSC (Poland); FCT (Portugal); MESTD (Serbia); MICIU/AEI and PCTI (Spain); MOSTR (Sri Lanka); Swiss Funding Agencies (Switzerland); MST (Taipei); MHESI and NSTDA (Thailand); TUBITAK and TENMAK (Turkey); NASU (Ukraine); STFC (United Kingdom); DOE and NSF (USA).

\hyphenation{Rachada-pisek} Individuals have received support from the Marie-Curie program and the European Research Council and Horizon 2020 Grant, contract Nos.\ 675440, 724704, 752730, 758316, 765710, 824093, 101115353, 101002207, 101001205, and COST Action CA16108 (European Union); the Leventis Foundation; the Alfred P.\ Sloan Foundation; the Alexander von Humboldt Foundation; the Science Committee, project no. 22rl-037 (Armenia); the Fonds pour la Formation \`a la Recherche dans l'Industrie et dans l'Agriculture (FRIA-Belgium); the Beijing Municipal Science \& Technology Commission, No. Z191100007219010, the Fundamental Research Funds for the Central Universities, the Ministry of Science and Technology of China under Grant No. 2023YFA1605804, and the Natural Science Foundation of China under Grant No. 12061141002 (China); the Ministry of Education, Youth and Sports (MEYS) of the Czech Republic; the Shota Rustaveli National Science Foundation, grant FR-22-985 (Georgia); the Deutsche Forschungsgemeinschaft (DFG), among others, under Germany's Excellence Strategy -- EXC 2121 ``Quantum Universe" -- 390833306, and under project number 400140256 - GRK2497; the Hellenic Foundation for Research and Innovation (HFRI), Project Number 2288 (Greece); the Hungarian Academy of Sciences, the New National Excellence Program - \'UNKP, the NKFIH research grants K 131991, K 133046, K 138136, K 143460, K 143477, K 146913, K 146914, K 147048, 2020-2.2.1-ED-2021-00181, TKP2021-NKTA-64, and 2021-4.1.2-NEMZ\_KI-2024-00036 (Hungary); the Council of Science and Industrial Research, India; ICSC -- National Research Center for High Performance Computing, Big Data and Quantum Computing and FAIR -- Future Artificial Intelligence Research, funded by the NextGenerationEU program (Italy); the Latvian Council of Science; the Ministry of Education and Science, project no. 2022/WK/14, and the National Science Center, contracts Opus 2021/41/B/ST2/01369 and 2021/43/B/ST2/01552 (Poland); the Funda\c{c}\~ao para a Ci\^encia e a Tecnologia, grant CEECIND/01334/2018 (Portugal); the National Priorities Research Program by Qatar National Research Fund; MICIU/AEI/10.13039/501100011033, ERDF/EU, "European Union NextGenerationEU/PRTR", and Programa Severo Ochoa del Principado de Asturias (Spain); the Chulalongkorn Academic into Its 2nd Century Project Advancement Project, and the National Science, Research and Innovation Fund via the Program Management Unit for Human Resources \& Institutional Development, Research and Innovation, grant B39G680009 (Thailand); the Kavli Foundation; the Nvidia Corporation; the SuperMicro Corporation; the Welch Foundation, contract C-1845; and the Weston Havens Foundation (USA).
\end{acknowledgments}

\bibliography{auto_generated}

\clearpage
\appendix
\numberwithin{figure}{section}
\numberwithin{table}{section}

\section{Additional Figures}
\label{app}

\begin{figure}[hbp]
  \centering
    \includegraphics[width=0.90\textwidth]{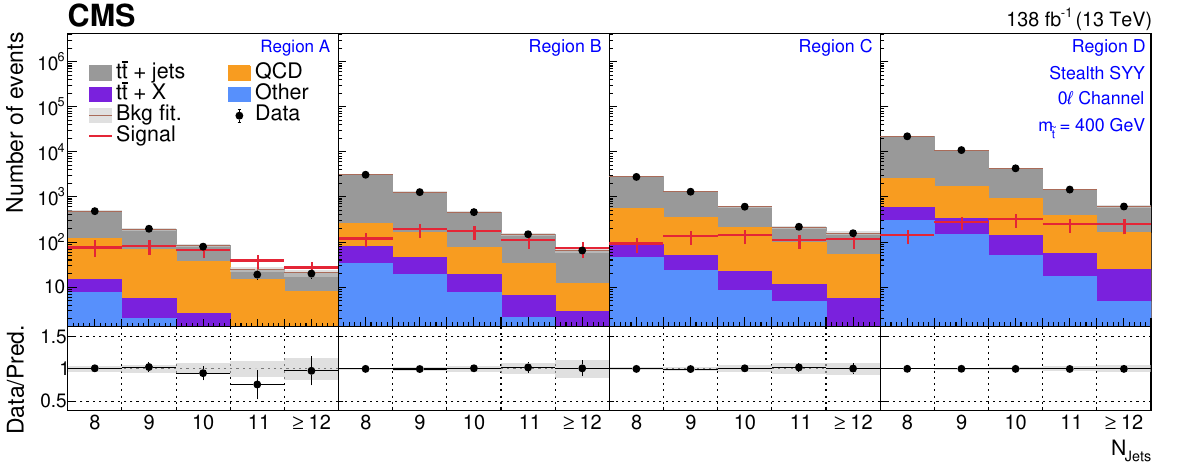}
    \includegraphics[width=0.90\textwidth]{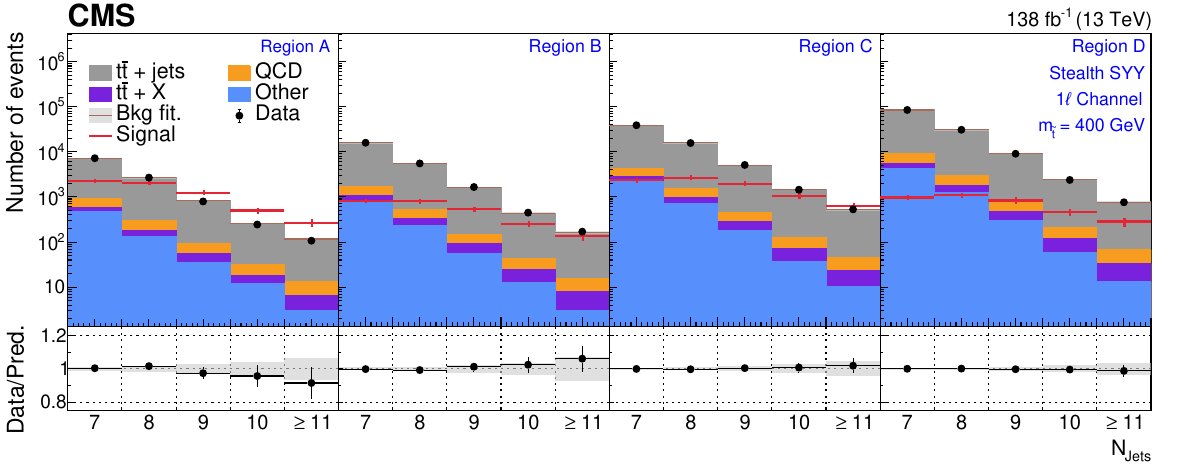}
    \includegraphics[width=0.90\textwidth]{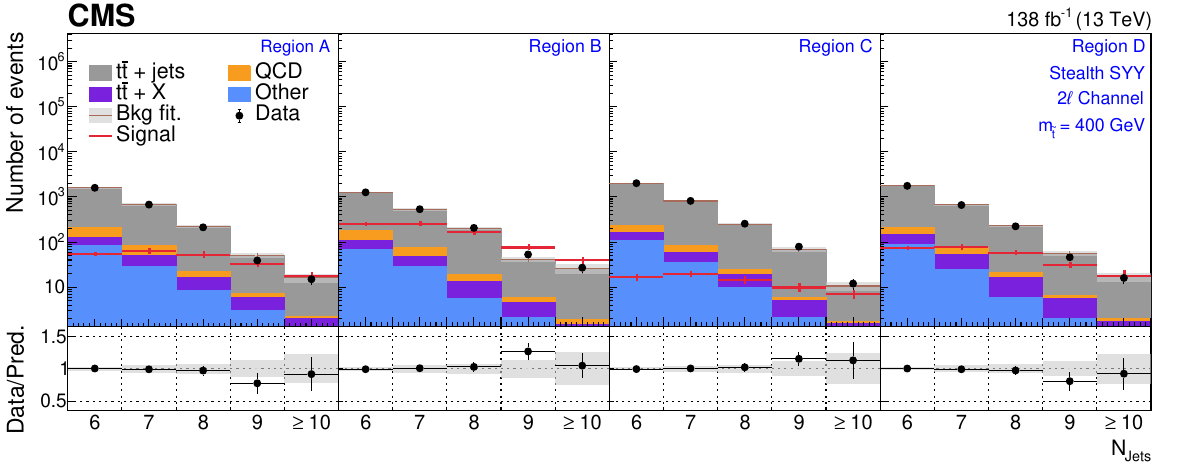}
  \caption{The \njets distributions from the background-only fits to data. Fits are run with the signal strength fixed to zero and all background process event yields are obtained from their fit predictions. The signal distribution overlaid corresponds to the stealth \syy model with $\mstop = 400\GeV$. The graphs in the upper, middle, and lower rows correspond to the \zerol, \onel, and \twol channels, respectively. The four panels in each row show the \njets distributions for the \Areg, \Breg, \Creg, and \Dreg regions (left to right). The gray error band shows the combined statistical and systematic uncertainty from the background-only fit distributions.}
  \label{fig:backgroundOnly_SYY_LowMass}
\end{figure}

\begin{figure}[p]
  \centering
    \includegraphics[width=0.90\textwidth]{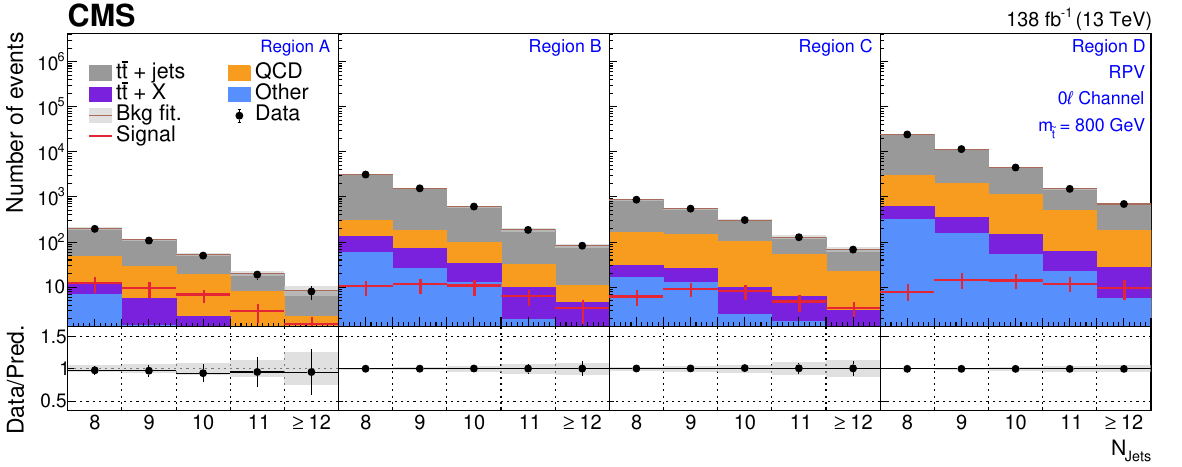}
    \includegraphics[width=0.90\textwidth]{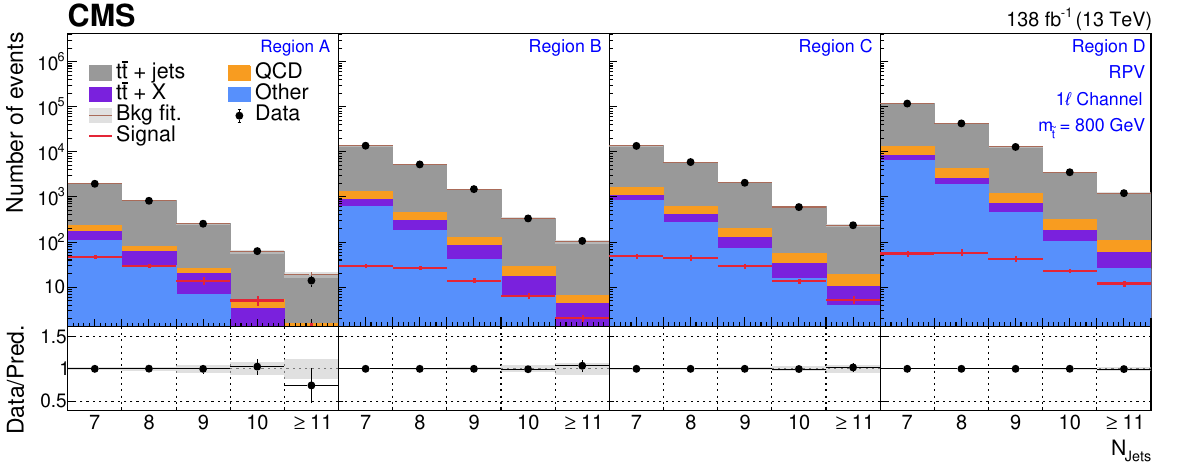}
    \includegraphics[width=0.90\textwidth]{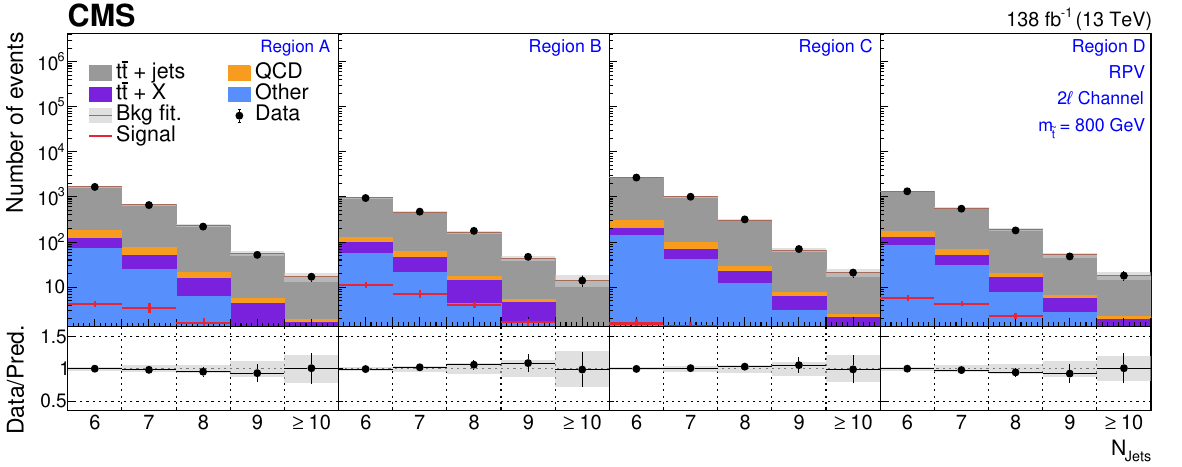}
  \caption{The \njets distributions from the background-only fits to data. Fits are run with the signal strength fixed to zero and all background process event yields are obtained from their fit predictions. The signal distribution overlaid corresponds to the RPV model with $\mstop = 800\GeV$. The graphs in the upper, middle, and lower rows correspond to the \zerol, \onel, and \twol channels, respectively. The four panels in each row show the \njets distributions for the \Areg, \Breg, \Creg, and \Dreg regions (left to right). The gray error band shows the combined statistical and systematic uncertainty from the background-only fit distributions.}
  \label{fig:backgroundOnly_RPV_HighMass}
\end{figure}

\begin{figure}[p]
  \centering
    \includegraphics[width=0.90\textwidth]{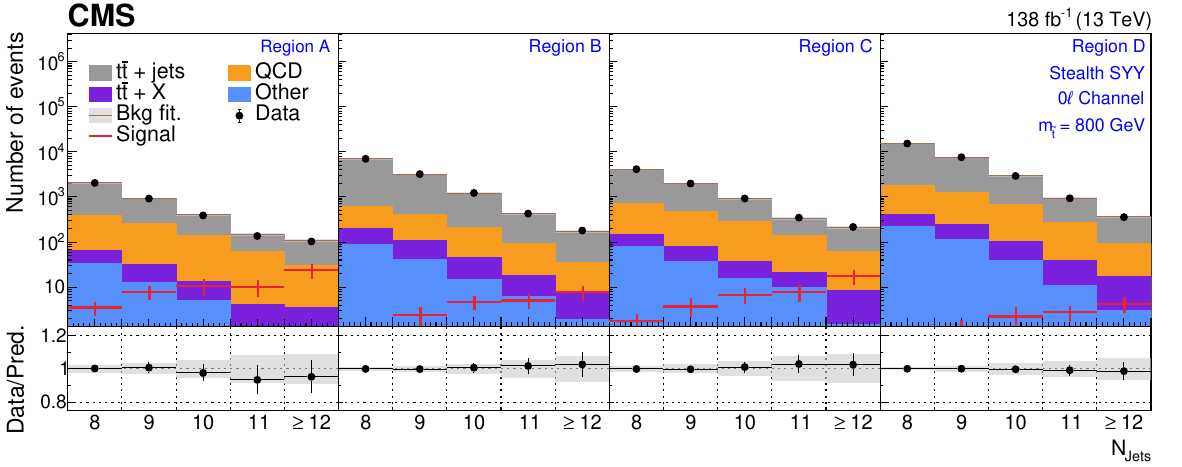}
    \includegraphics[width=0.90\textwidth]{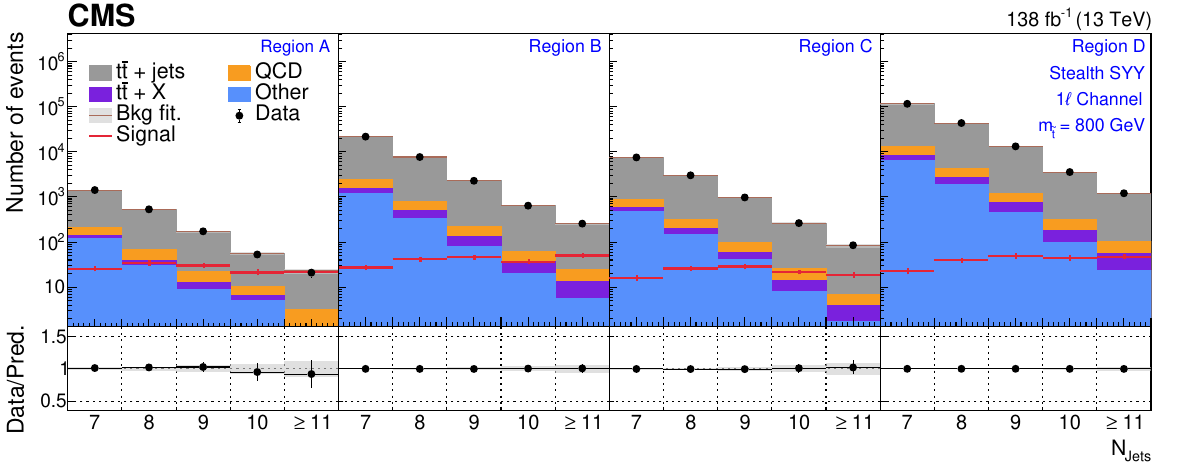}
    \includegraphics[width=0.90\textwidth]{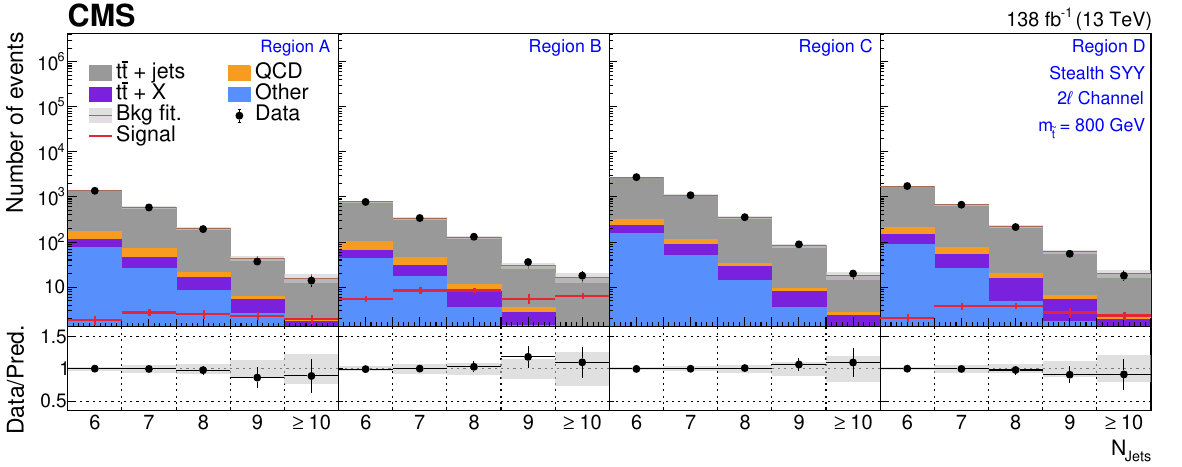}
  \caption{The \njets distributions from the background-only fits to data. Fits are run with the signal strength fixed to zero and all background process event yields are obtained from their fit predictions. The signal distribution overlaid corresponds to the stealth \syy model with $\mstop = 800\GeV$. The graphs in the upper, middle, and lower rows correspond to the \zerol, \onel, and \twol channels, respectively. The four panels in each row show the \njets distributions for the \Areg, \Breg, \Creg, and \Dreg regions (left to right). The gray error band shows the combined statistical and systematic uncertainty from the background-only fit distributions.}
  \label{fig:backgroundOnly_SYY_HighMass}
\end{figure}

The background-only fit distributions shown in the main body of this work show the results 
of the analysis for the RPV signal model using the low-mass optimization. 
Figures~\ref{fig:backgroundOnly_SYY_LowMass}, \ref{fig:backgroundOnly_RPV_HighMass}, and \ref{fig:backgroundOnly_SYY_HighMass} present the background-only
fit results for the stealth \syy analysis using the low-mass optimization as well as the background-only fits for both signal topologies using the high-mass optimization.

\cleardoublepage \section{The CMS Collaboration \label{app:collab}}\begin{sloppypar}\hyphenpenalty=5000\widowpenalty=500\clubpenalty=5000
\cmsinstitute{Yerevan Physics Institute, Yerevan, Armenia}
{\tolerance=6000
V.~Chekhovsky, A.~Hayrapetyan, V.~Makarenko\cmsorcid{0000-0002-8406-8605}, A.~Tumasyan\cmsAuthorMark{1}\cmsorcid{0009-0000-0684-6742}
\par}
\cmsinstitute{Institut f\"{u}r Hochenergiephysik, Vienna, Austria}
{\tolerance=6000
W.~Adam\cmsorcid{0000-0001-9099-4341}, J.W.~Andrejkovic, L.~Benato\cmsorcid{0000-0001-5135-7489}, T.~Bergauer\cmsorcid{0000-0002-5786-0293}, S.~Chatterjee\cmsorcid{0000-0003-2660-0349}, K.~Damanakis\cmsorcid{0000-0001-5389-2872}, M.~Dragicevic\cmsorcid{0000-0003-1967-6783}, P.S.~Hussain\cmsorcid{0000-0002-4825-5278}, M.~Jeitler\cmsAuthorMark{2}\cmsorcid{0000-0002-5141-9560}, N.~Krammer\cmsorcid{0000-0002-0548-0985}, A.~Li\cmsorcid{0000-0002-4547-116X}, D.~Liko\cmsorcid{0000-0002-3380-473X}, I.~Mikulec\cmsorcid{0000-0003-0385-2746}, J.~Schieck\cmsAuthorMark{2}\cmsorcid{0000-0002-1058-8093}, R.~Sch\"{o}fbeck\cmsAuthorMark{2}\cmsorcid{0000-0002-2332-8784}, D.~Schwarz\cmsorcid{0000-0002-3821-7331}, M.~Sonawane\cmsorcid{0000-0003-0510-7010}, W.~Waltenberger\cmsorcid{0000-0002-6215-7228}, C.-E.~Wulz\cmsAuthorMark{2}\cmsorcid{0000-0001-9226-5812}
% ...
\par}
\vskip\cmsinstskip
\dag:~Deceased\\
$^{1}$Also at Yerevan State University, Yerevan, Armenia\\
$^{2}$Also at TU Wien, Vienna, Austria\\
% ...
\end{sloppypar}
\end{document}